\documentclass[12pt,preprint]{aastex}
\usepackage{graphicx}

\shorttitle{Local Environment of ULXs with OM}
\shortauthors{Berghea et al.}

\begin{document}
\title{The Local Environment of Ultraluminous X-ray Sources Viewed by {\it XMM Newton}'s Optical Monitor}

\author{C. T. Berghea and R. P. Dudik} 
\affil{United States Naval Observatory, Washington, DC 20392}
\email{ciprian.berghea@usno.navy.mil}
\email{rachel.dudik@usno.navy.mil}

\author{J. Tincher}
\affil{St. John's College, Annapolis, MD 21401}

\author{L. M. Winter}
\affil{Atmospheric and Environmental Research, Lexington, MA 02421}

\begin{abstract}
We have used {\it XMM-Newton's Optical Monitor} (OM) images to study the local environment
of a sample of 27 Ultraluminous X-ray Sources (ULXs) in nearby galaxies. 
UVW1 fluxes were extracted from 100~pc regions centered on the ULX positions.  We find that at least 4 ULXs (out of 10 published) have spectral types that are consistent with previous literature values.  
In addition the colors are similar to those of young stars.
For the highest-luminosity ULXs, the UVW1 fluxes may have an important contribution from the accretion disk.
We find that the majority of ULXs are associated with recent star-formation.
Many of the ULXs in our sample are located inside young OB associations or star-forming regions (SFRs).
Based on their colors, we estimated ages and masses for star-forming regions located within 1~kpc from the ULXs in our sample.
The resolution of the OM was insufficient to detect young dense super-clusters, but some of these star-forming regions 
are massive enough to contain such clusters. Only three ULXs have no associated SFRs younger than $\sim$50~Myr.
The age and mass estimates for clusters were used to test runaway scenarios. 
The data are in general compatible with stellar-mass binaries accreting at super-Eddington rates and ejected by natal kicks.
We also tested the hypothesis that ULXs are sub-Eddington accreting IMBHs ejected by three-body interactions,
however this is not supported well by the data.

\end{abstract}

\keywords{X-rays:binaries - ultraviolet: stars - galaxies: star clusters: general - accretion, accretion disks}

\section{INTRODUCTION}

Ultraluminous X-ray Sources - extremely bright X-ray sources with bolometric luminosities 
exceeding the Eddington limit for a 20\,M$_{\sun}$ object - continue to puzzle astronomers.
Due to their distance, located in external galaxies, it is difficult to identify optical counterparts,
even with the Hubble Space Telescope \citep{ptak06,ram06, rob08,tao11,glad13}. Indeed very few objects have been detected in the optical with high confidence $-${\bf  this is compared with the hundreds detected in the X-rays}\footnote{{\bf The following is a list of references for some of the most famous objects with optical counterparts:  NGC~5204~X-1 \citep{liu04}, NGC~1313~X-1 \citep{yang11}, ESO~243-49 HLX-1 \citep{sor12}
NGC~1313~X-2 \citep{zam04,liu07,rip11,zam12}, M81 X-6 \citep{liu02b,swa03,moon11}, Holmberg~IX X-1 \citep{gris06,moon11,gris11},
Holmberg~II~X-1 \citep{kaa04,tao12b}, NGC~5408 X-1 \citep{lang07,gris12}, M101 X-1 \citep{kun05,liu09M101}, NGC~4559 \citep{sor05},
two ULXs in M51 \citep{ter06}, NGC~2403 X-1 \citep{rob08}, IC~342 X-1 \citep{feng08}, the ULX in NGC~247 \citep{tao12a},
NGC 6946 X-1 \citep{kaa10} and ULX P13 in NGC 7793 \citep{pak10,motch11}}.}.  In addition, only three objects have established optical periods that have been measured: M82 \citet{kaa06,kaa07}, for NGC~1313 X-2 \citet{liu09,zam12} and NGC~5408 X-1 \citep{stro09,han12}.

{\bf Multiple authors have shown that ULXs are associated with star formation in their host galaxies \citep{ran03, gri03}.}  Indeed such a correlation has been observed for high-mass X-ray binaries (HMXBs) in our own Galaxy \citep{gri03}. 
\citet{bod12} found an average offset of $\sim$400~pc between galactic HMXBs and nearby SFRs \citep[see also][]{col13}.
Many authors have found evidence that ULXs are generally associated with large clusters and that the ULX is nearby these clusters.  
For instance \citet{kaa04} found that in starburst galaxies, X-ray sources are, in general, located near star clusters
but also that X-ray sources with luminosities $>$~10$^{38.0}$~erg~s$^{-1}$ tend to be even closer to clusters.
Similar results were found by \citet{ran11} in the starburst galaxy NGC~4449.  Eleven X-ray binaries were found to be located nearby or inside very young clusters.
In the Antennae galaxy, 10 out of 14 ULXs seem to be associated with young stellar clusters \citep{zez02}.
\citet{cla07} found 7 ULXs associated with clusters in the Antennae galaxy using infrared images. 
These authors found that, in general, X-ray sources tend to be close to large clusters \citep[see also][]{clark11}.
\citet{swa09} used photometric data from the Sloan Digital Sky Survey to look for possible associations of 47 ULXs with SFRs or young superclusters.
They found that statistically ULXs are indeed associated with recent star formation (within 100~pc distance), but no superclusters were detected given the poor spatial resolution of the instrument.
\citet{pou12} performed spectral and photometric analyses of clusters associated with the Antennae ULXs
and found that almost all are very young (2.4 to 3.2 Myr), and that only one resides inside a cluster \citep[see also][]{ran12}.\
Originally M82 X-1 was thought to be located inside a supercluster, until \citet{voss11} 
showed that it is actually offset from the cluster.  However, in the same paper, the authors found that a ULX in NGC~7479 was associated with a young supercluster, so such objects are known to exist.

{\bf In theory the location of a ULX in relation to its surrounding star clusters can tell something about the environment in which the black hole was born as well as constrain some of the properties of the black hole.  For instance, if a small black hole is born in a cluster of stars, the initial explosion can be asymmetric enough to kick the black hole out of the cluster.  This is known as the runaway binary scenario \citep{zefa02}.  This theoretical scenario does not work for larger black holes, such as intermediate black holes \citep[IMBH][]{col99}, since a) the black hole is too big to be susceptible to such kicks and b) even if such a black hole were kicked it would return to the cluster on very short timescales because of gravitational pull \citep[ i.e. v$_{kick}$ $<$ v$_{escape}$]{por99}.   However, there are ways to kick intermediate black holes out of clusters using 3-body interactions \citep[e.g.][]{pou12}.  In this scenario the intermediate mass black hole and donor star can be kicked out of the cluster by another young massive interloper star.  Assuming this theory is correct, we would expect to find most potential intermediate mass black holes inside young, dense clusters of stars.   Such, an environment would also readily explain the growth of an intermediate mass black hole through stellar collisions with the stars in the parent cluster \citep[e.g.][]{gur06}.  Thus, in summary, all of this information can be combined to estimate characteristics of the ULX including the age and mass of a stellar companion assuming the black hole originated in a nearby cluster of stars or parent cluster \citep{zefa02,kaa04}.}

The observational evidence for a possible association between ULXs and the star-formation in the galaxy is still controversial.
Some starburst galaxies such as the Antennae galaxy, the Cartwheel galaxy and M82 
contain an unusually large number of very bright ULXs.  On the other hand,
there are many star-forming galaxies without any known ULXs.  In addition there are dwarf galaxies (such as Holmberg~II and Holmberg~IX) and even some ellipticals that contain bright ULXs.
The ULXs found in elliptical galaxies seem to be fainter.  Some of the bright ULXs that appear may actually be interlopers \citep{irw04}.  Interestingly, most of the galaxies mentioned here (Antennae, Cartwheel, M82, Holmberg II, and Holmberg IX) are either merging or interacting with other galaxies.   More recent examples are the colliding galaxy pair NGC~2207/IC~2163 where 21 ULXs were detected \citep{min13}, Arp~147 with 9 ULXs \citep{rap10}, and NGC~922 with 12 ULXs \citep{pre12}. The latter two are drop-through ring galaxies, similar to the Carthwheel.

In this paper we use {\it XMM-Newton} archival images taken with the Optical Monitor (OM) 
to explore ULX environments in nearby galaxies.
Ultraviolet (UV) emission is well suited to study star-forming regions (SFRs) and young clusters around ULXs.
The ULX sample presented here was extracted from the XMM sources analyzed in \citet[][hereafter, WMR]{win06}.
We selected all the ULXs with unabsorbed luminosities L$_X$~$\ge$~2.7$\times$10$^{39}$~erg~s$^{-1}$, 
as estimated by the authors. This limit corresponds to a 20~M$_{\odot}$ black hole radiating at the Eddington limit.
Our ULX sample consists of all the sources for which OM UV data was found in the archives.
We added to this sample the ULX associated with the MF16 nebula in NGC~6946 as a standard, since it is well-studied in the literature.
The original ULX sample in WMR was selected from galaxies closer than 8~Mpc.
The goal was first to examine whether ULXs are located inside clusters or SFRs 
or are possibly related to nearby such regions.  Second was to use photometry to impose constraints
on the nature of the optical companions and the accretion mechanism of the ULX.
We measured fluxes for 100~pc regions at ULX positions and for bright sources (SFRs) detected nearby.
Section~2 describes the data found in the archives and the photometry procedure.
We present the photometry results for the 100~pc region centered on ULXs in Section~3.
The results for the SFRs found close to ULXs are presented in Section~4,
together with the population synthesis modeling used to estimate ages and masses.
These are used in Section~5 to test runaway binary scenarios.
Finally, in Sections~6 and 7 we discuss our results and present the main conclusions.

\section{OBSERVATIONS AND DATA REDUCTION}

OM is a 30-cm optical/UV instrument co-aligned with the X-ray instrument on the {\it XMM-Newton} telescope.
The OM includes three optical and three UV filters: V, B, U, UVW1, UVM2, and UVW2, with effective
wavelengths of 5430\AA, 4500\AA, 3440\AA, 2910\AA, 2310\AA, and  2120\AA, respectively.
The OM instrument has a smaller field of view than the X-ray instrument, but was not used by default in all observations. We found 26 out of 36 ULXs from WMR that had been observed by the OM and which met our criteria for selection (ULX is defined as any object with L$_X$~$\ge$~2.7$\times$10$^{39}$~erg~s$^{-1}$, see Section 1  for details).  We also included the ULX in NGC~6946 as a reference object, because it is well studied in the literature.  All 27 objects  are located in 16 nearby galaxies. 
We found a total of 36 useful OM observations of these galaxies in the archives as of 2012 Jun 25 (meaning that some ULXs have more than one observation). 
These are listed in Table~\ref{table1}. 
The ULXs were observed with a diverse set of filters.  Most ULXs were not observed with all of the filters, however. 
We discarded images with artifacts like ``ghost images''.
However, with the exception of NGC 4736 XMM1 and NGC 5408 XMM1, 
all ULXs have images with the most sensitive of the UV filters, UVW1.
For the large majority of ULXs we found images in at least three filters, 
allowing the construction of color-color plots.  When redundant images were available we used those images without artifacts and with longer exposures.

We used the ``sky images,'' which are processed images created by the OM pipeline.  We used the {\it XIMAGE} software included in the {\it HEAsoft v6.12} package to measure the count rates in each filter.   We converted the count rates into flux values
using the count rate to flux conversion factors listed in the {\it XMM-Newton Science Analysis System Users Guide}\footnote{http://xmm.esac.esa.int/external/xmm\_user\_support/documentation/sas\_usg/USG/}.
We applied the coincidence-loss and dead-time corrections on each pixel of the extended source, 
following the procedure  described in the {\it XMM-Newton Users Handbook}\footnote{http://xmm.esac.esa.int/external/xmm\_user\_support/documentation/uhb/}.

\begin{figure}
\rotate
\epsscale{0.9}
\plottwo{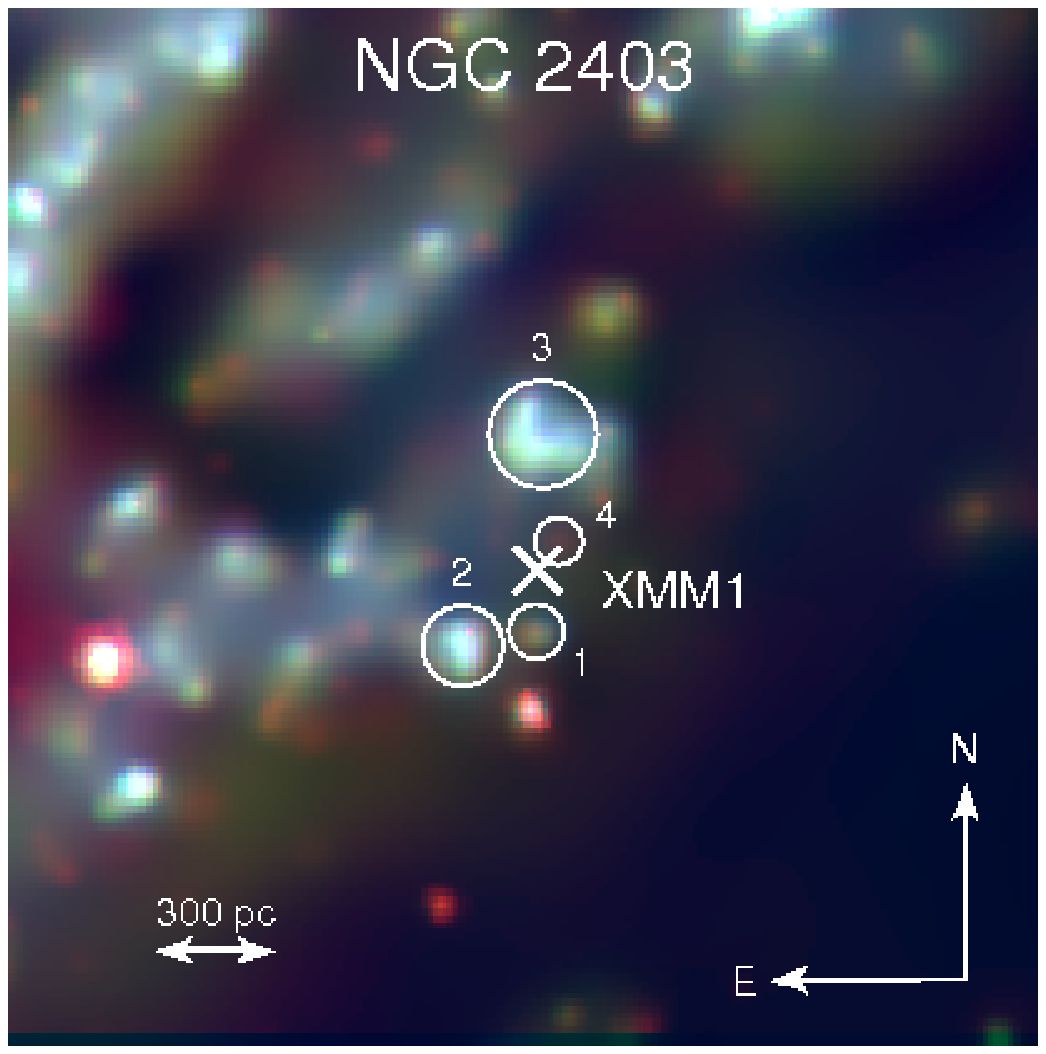}{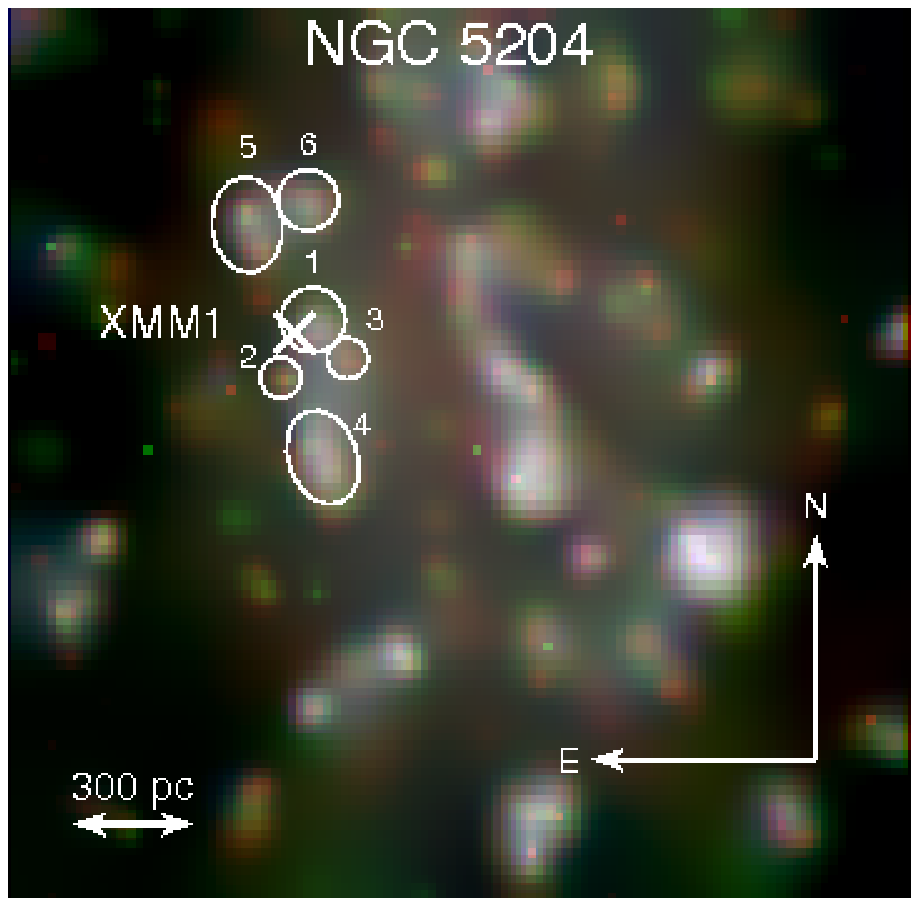}
\caption{
Color coded and adaptively smoothed OM UV images of NGC 2403 and NGC 5204, showing the extraction regions of ULXs
and nearby star forming regions. {\bf The extraction region around the ULX position is 100pc for all ULXs, but the extraction region around the surrounding star forming regions varies in size based on the extent of the emitting region (see Section~2 for details).} The ``X'' marks the ULX position.  Colors are for the three UV filters on OM:
UVW1 - red, UVM2 - green and UVW2 -  blue.
} \label{images}
\end{figure}

{\bf {\bf ULX 100~pc regions:}  To do the astrometry, all of the OM images were taken from \citet{kun08}, who derive coordinates for each image based on the USNO-B1 catalog which is tied to the International Celestial Reference Frame (ICRF).   \citet{kun08} estimates the errors on these positions to be 0.3$\arcsec$.  When an optical counterpart was observed, small shifts in the images were made to align the OM and {\it Chandra} images further, thereby deriving the absolute position of the ULX to less than the 0.3$\arcsec$ accuracy of the \citet{kun08} images . 

For each source, we created regions centered on the ULX position with diameters of  100~pc, assuming the distances listed in Table~1 of WMR (two examples are shown in Figure~\ref{images}). The 100~pc regions translate into angular sizes ranging from 2.6$\arcsec$ to 6.6$\arcsec$. The 100~pc regions are the smallest possible regions that could be extracted for photometry since the FWHM of the point spread function (PSF) for the OM instrument ranges from 1.5$\arcsec$ to 2.0$\arcsec$.  The only exception to the 100~pc extraction region was IC~342 XMM3 which is located inside a very UV-bright nuclear cluster.  For this ULX we used a region of 265~pc diameter instead of 100~pc, in order to include the extended emission.}

{\bf {\bf Surrounding Star Forming Regions:}  For each ULX we also measured fluxes from nearby SFRs, using circular or elliptical apertures (of varying sizes, with offsets up to 1~kpc from each ULX). The number of these regions, their sizes, and offset distances vary for each object. For most ULXs we measured fluxes for 3$-$5 nearby SFRs (see Figure~\ref{images} for two examples).
In general, they are large star-forming regions with diameters of up to 500~pc.  The regions are almost always brighter than the 100~pc regions around the ULX  and the fluxes measured are more accurate as a result. Aligning the sources between filters is straight forward, since there are multiple stellar sources with which to do the relative astrometry between filters.}  

\begin{figure}
\rotate
\epsscale{0.7}
\plotone{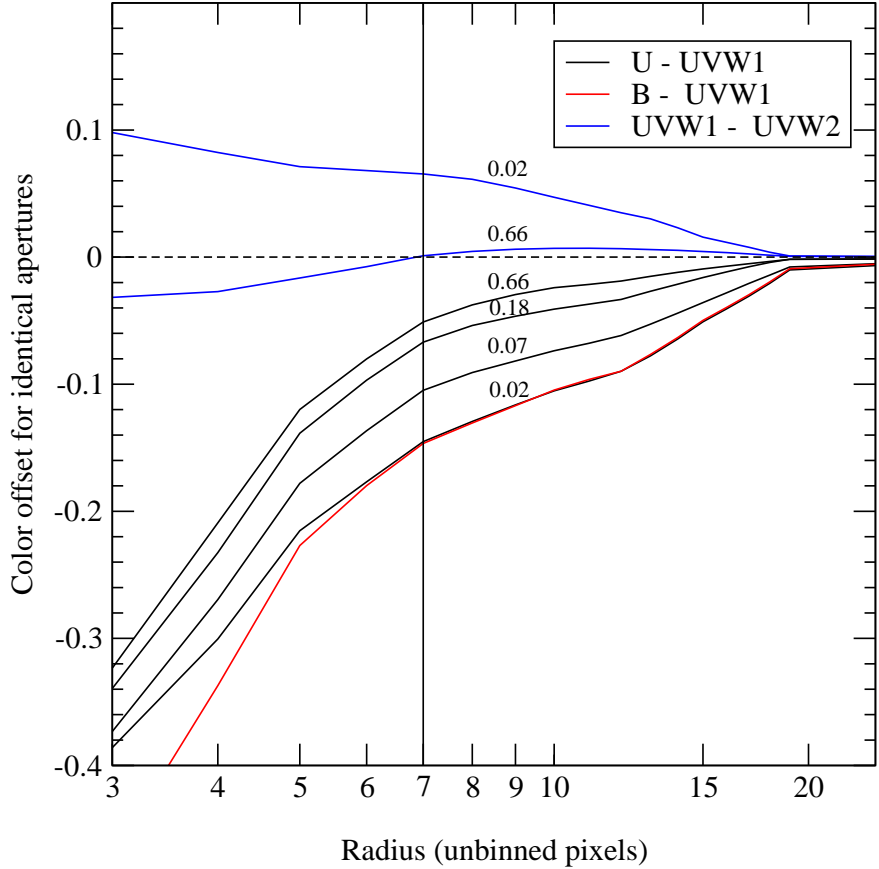}
\caption{
Magnitude offsets caused by using the same sized aperture for each region across all OM filters. The black lines are U$-$UVW1 color offsets versus aperture size, for point sources with various count rates.  The count rates are shown above each line.  The vertical line shows the radius of the aperture we used for many of our sources as an example.  For faint sources we used a maximum offset U$-$UVW1~$=-$0.15.  For brighter sources we interpolated this value down to the minimum U$-$UVW1~$=-$0.06. The B$-$UVW1 color offsets are shown (in red) and are very similar to U$-$UVW1.  Finally, for UVW1$-$UVW2 (in blue), the offsets are smaller but still significant.   In this case we used a maximum offset of UVW1$-$UVW2~$=$~0.06 for faint sources and zero offset for the brightest sources.'
} \label{psf}
\end{figure}

{\bf {\bf Errors and Corrections:}  Most of the UV light from the ULXs is extended (i.e. not a point source) and most of the emission is very faint.   Because of this, the resultant flux from the 100~pc regions could include multiple sources rather than a single source that could easily be fit with the OM point source extraction software.  This leads to confusion in the measurement for the 100pc region around the ULX.

Additional errors are introduced through aperture effects for both the ULX 100pc regions and the chosen star forming regions.  These effects result from using the same-sized apertures for each extraction region across all wavebands.  These color-based aperture corrections are complicated by the extended emission (i.e. that the emission is not from a point source).  Indeed the needed corrections are higher for colors calculated between the optical and the UV filters, since the point spread functions (PSFs) between these filters are different.

In Figure~\ref{psf} we calculate the theoretical color offsets obtained when using apertures of the same sizes.  We use the latest PSF calibration files available from OM\footnote{ftp://xmm.esac.esa.int/pub/ccf/constituents/OM\_PSF1DRB\_0010.CCF}.  The photometry for OM is calibrated for extraction apertures with radii of 12 and 35 pixels, for the optical and UV, respectively.  The OM PSFs vary with source brightness or, in other words, the ratio of the count rate to frame rate.
The FWHM decreases with source brightness, and therefore the offsets for the plotted U - UVW1 color also decrease.   The offsets between optical and UV filters are similar, as illustrated by the B - UVW1 color in red (compared with U - UVW1). The color offsets between the various optical filters are very small because the PSFs are quite similar, however there is a slight offset between the UVW1 and UVW2 filters, which is much smaller than the optical to UV offsets, but which is large enough to warrant correction (blue lines). }

Based on the theoretical results shown in Figure~\ref{psf}, we introduced color corrections.  However we note that, since practically all of the sources in this paper are extended, the correction is approximate. 
We corrected the measured colors between the optical and UV filters (U-UVW1 and B-UVW1) by interpolating on the source flux up to $-$0.15 mag for faint sources (see also Figure~\ref{psf}). For the UVW1 - UVW2 color, 
we used a maximum offset of 0.06 mag for faint sources, scaling to zero for the bright sources. 
We found that these corrections give consistent estimates for the cluster ages when other colors are available to make the same calculation (e.g. M101 which has been observed in all filters). In general these corrections are not large and are in fact approximately the same size as the measurement errors on the original magnitudes.

All extraction regions (both 100~pc ULX regions and SFRs) were background corrected.  The background  was measured by extracting fluxes from nearby areas that were free of visible point sources.

The color excess values E(B-V) for the Milky Way extinction were obtained
from the NASA Extragalactic Database (NED). Both Galactic and extragalactic extinctions
were calculated using the extinction curves from \citet{car89}, assuming R$_V=$~3.1.  {\bf All photometric magnitudes presented were corrected using Galactic extinctions and the intrinsic extinctions obtained from the color-color models presented in Section~4.}

\section{PHOTOMETRY RESULTS FOR THE 100 PC REGIONS}

In this Section we use the fluxes measured from the 100~pc regions around the ULX to check for possible UV counterparts. These could be single stars (the companions in the ULX binaries),
small OB associations or larger star-forming regions (clusters). 
An important contribution to the optical or UV flux could also come from the accretion disk.
We first use the UVW1 fluxes to make general estimates for the different possible counterparts.
Then, where available, we use the colors in different OM bands to compare with models of 
standard stars and accretion disks.

\subsection{UVW1 Magnitudes}

The UVW1 is the most sensitive of the UV filters and most of our ULXs have images taken with this filter.  NGC 4736 XMM1 and NGC 5408 XMM1 have no images in the UVW1 band.  For these ULXs we used the U and UVM2 band, respectively.  {\bf For ULXs with multiple observations we used the error-weighted average.  Based on the {\it XMM-Newton Users Handbook}, we estimate the detection limit for a O5 V star to be $\sim$18.8 mag in the UVW1 filter.  This means an individual O5 V star will be detected to about 2.4~Mpc with the UVW1 filter if the exposure time is 1000~s.  Our ability to measure single stars around ULXs is difficult because the brightness of these stars is very close to the detection limit of the UVW1 filter, though the average exposure time is longer (3000~s).}  The magnitudes provided in Table~\ref{table2} should therefore be treated as upper limits.  The measured UVW1 magnitudes are later used to estimate the nature of the UV emission (i.e. whether the flux is consistent with values expected from a single star, multiple stars, or accretion emission).

\subsubsection{Single stars}
{\bf In Column~11 of Table~\ref{table2}, we list the estimated stellar type based on the UVW1 magnitudes, thereby assuming that all the flux within the 100~pc region is emitted by a single star.  Given the signal-to-noise (S/N) ratios in Column~9 of Table~\ref{table2} (we take S/N~$>$~3 for a detection), we estimate that 7 ULXs {\it may} have single stars as counterparts.}  An additional two ULXs have no significant detections.  For these we show 3$\sigma$ upper limits.  In Column~12 of Table~\ref{table2} we also show the stellar types for all previously identified optical counterparts in the literature.   {\bf We note that while the absolute magnitude is consistent with that of a single star, the UV emission from these objects are also likely contaminated by accretion disk emission as has been shown in many cases \citep{roberts2011, van1994}.}

{\bf Ten of our objects have estimated spectral types for donor stars in the literature.   Of these four ULXs (NGC~1313 XMM3, NGC~2403 XMM1, Holmberg~IX XMM1 and M101~XMM3) have UVW1 spectral-type predictions that are identical or very close to that found in the literature.} For the other six ULXs which have previous optical counterparts in the literature, the 100~pc UV fluxes suggests multiple stars. In these cases the ULXs are located inside or very close to small clusters or OB associations and therefore the UV fluxes are likely contaminated by surrounding star formation. The following outlines background information for ULXs which may suffer contamination:

{\bf IC~342 XMM1:} IC~342 XMM1 has a F8-G0~Ib supergiant as an optical counterpart according to \citet{feng08}.  They find no O stars within 300~pc of the ULX, and estimate that for detected stars, the minimum stellar age is 10~Myr.
If this star is indeed a G0, then the absolute magnitude of the star is only $-$3.1, and is well below the detection threshold of the UVW1. We estimate that the UVW1 flux is equivalent to five O5V stars, but given the previously estimated age, the UV emission is likely produced by several later type stars. 

{\bf Holmberg~II XMM1:} Homberg~II XMM1 (HoII~X-1) has a 21.5 V-band magnitude stellar counterpart with colors consistent with O stars on the main sequence or B supergiants \citep{kaa04}.  The ULX is located inside an association of young stars \citep{kaa04}.  We suspect that here too, the total UVW1 flux is overestimated due to contamination.  

{\bf NGC~5204 XMM1:}  NGC~5204 XMM1 (better known as X-1) has a B0~Ib supergiant counterpart, 
based on the Si~III $\lambda$1299 line \citep{liu04}. The Hubble images in \citet{liu04}
show at least 3 star clusters and OB associations within our 100~pc region (4.3$\arcsec$). 
We estimate approximately eight O stars from our UVW1 flux, though here too the flux is likely contaminated. 

{\bf NGC~5408 XMM1:}  NGC~5408 XMM1 is also located very close to an OB association.
\citet{gris12} estimated its age at 5~Myr and also identified a B0 supergiant 
as the likely ULX counterpart. Several young stars from the OB association
fall into our 100~pc extraction region.  Instead of a single star, we find  the UVW1 flux is equivalent to 8 O5V stars.

{\bf {\bf NGC~247 XMM1:} \citet{tao12a} used HST observations to show that the ULX is very close to a large and relatively young stellar association. The stars closest to the ULX are contaminating the UVW1 flux.  The estimate is therefore equivalent to 5 O5V stars.

{\bf M51~XMM6:} The ULX is located at the edge of a large SFR, its emission dominates our 100~pc flux.  The measured UVW1 flux is equivalent to 35 O5V stars. 

{\bf NGC~6946 XMM1:} This ULX is associated with the bubble nebula MF16 and a UV point source was identified using HST data by \citet{kaa10}. Several nearby bright stars and the nebular emission are contaminating our UVW1 flux here as well \citep[see also][]{ber12}.  The measured UVW1 flux is equivalent to $\sim$ 3 O5V stars.}

Finally, for three ULXs (not including the upper limits) we have no previous published information.  For these, even if our estimates are compatible with single stars, the UV images suggest extended star-forming regions, even beyond the 100~pc regions. The UVW1 emission here is more likely produced by many late-type stars, instead of a single young star, given the spatial information we have.

\subsubsection{Multiple stars or star-forming regions}

For 18 ULXs we measured UVW1 fluxes that are too high for single stars,
they seem to be located inside small OB associations or SFRs. 
For these, Column~11 of Table~\ref{table2} lists the equivalent number of O5V stars,
assuming that all the UVW1 flux inside the 100~pc regions comes from such stars.  Two of the most obvious examples are IC~342 XMM3 and NGC~4736 XMM1.  

{\it IC~342 XMM3:}  IC~342 XMM3 is very close to the nucleus of the galaxy. 
\citet{kong03} estimated that it is located 3$\arcsec$ from the dynamical center of the galaxy.
\citet{bok99} measured the nuclear cluster in IC~342 using infrared and optical observations and found
a mass $<$~6$\times$10$^6$~M$_{\sun}$, and a relatively young age (10$^{6.8-7.8}$ yr), 
but found no evidence for a supermassive black hole.  If this age estimate is correct, O-stars are too young to be found in the region.  Indeed, using the models of \citet{lei99} we find O-stars in the number suggested will produce the emission observed in less than 1~Myr.  By 10$^{6.8-7.8}$ years \citep[the lower limit from][]{bok99}, all of these O-stars would be dead.

{\bf {\it NGC~4736 XMM1:}  For NGC~4736 XMM1 we estimate 747 O5V stars. The ULX is located inside the face-on spiral galaxy NGC~4736. HST images show strong unresolved star formation at the ULX location, and our estimated extinction is also high (Table~\ref{table3}).}

\subsubsection{Contributions from accretion disks}
In Fig.~\ref{bh} we plot the UVW1 magnitudes against the bolometric luminosities for all of the ULXs in our sample (bolometric luminosities were taken from WMR). The diagonal line in the upper-right corner of this image represents the expected accretion disk emission in the UV as a function of the bolometric magnitude.  This is a theoretical estimate, calculated using a standard disk model for a ULX accreting at the Eddington limit \citep{shak73}.  {\bf The dotted line represents the emission from an irradiated accretion disk, which is also the maximum contribution to the UV emission we would expect from a disk \citep{san93, ber10a}.}

The position of this line implies that a significant portion of the measured UV flux is likely from accretion disk emission if the ULX has a bolometric luminosity in excess of 10$^{40.0}$~erg~s$^{-1}$. {\bf This is also likely the case for sources appearing close to the theoretical model in the figure, such as NGC~1313 XMM3 and Holmberg~IX XMM1 (see Fig.~\ref{bh}). Interestingly these results are consistent with the findings of \citet{roberts2011}, who suggest, based on optical spectroscopy, that some portion of the UV emission from NGC~1313 XMM3 and Holmberg~IX XMM1 comes from a disk component.  We note that these ULXs also show some of the highest X/UVW1 ratios in Column~10 of Table~\ref{table2}.}

An even larger fraction of the UV emission would be expected to come from the disk if it were irradiated.  In fact, the UV emission can be increased by up to a factor of 10 for stellar mass BHs as shown in the Figure \citep{ber10a}.  If this is the case for NGC~1313 XMM3 and Holberg~IX XMM1, it would imply that the UVW1 flux is almost entirely from the disk.  

We also note that, unlike typical AGN, we do not see a trend in X/UVW1 (where X is a proxy for the bolometric luminosity).  Indeed there is no correlation at all for either single star sources or multi-star sources. 

{\bf While we cannot rule out a strictly stellar origin for the ULXs in our sample, it is very likely that a fraction of the UV emission we observe is from the accretion disk.  However as Fig.~\ref{bh} shows, for most of our sources, the accretion disk is not bright enough to be responsible for {\it all} of the UV emission (even if it is irradiated).  As mentioned previously (and as Table~\ref{table2}, column~11 shows) most of our sources are bright enough that the emission is likely to be contaminated by multiple stellar sources within the 100~pc aperture, thereby making it impossible to separate the two.  In addition, for those sources with stellar estimates, the UV emission is bright enough to come from both the donor star and a disk.  Since one of the main goals of this paper is to qualify the environments around the ULX, we feel that the approximate spectral type is very important to understanding the origin and age of the environment from which the ULX was born.  Again, this does not mean that some of the total emission does not come from the accretion disk, but that the photometry looks like a specific type of star. } 

\subsubsection{UVW1 variability}

{\bf Some of the objects in this sample have more than one observation taken in the UVW1 band, however only 4 objects have more than one observation in UVW1 where the source was detected (many of our objects have multiple observations but the others are non-detections).   We therefore looked into the variability of these four objects.  The results of the analysis are shown in Table~\ref{table4}.   Of the four objects, three (IC~342 XMM1, M~101 XMM2, and NGC~1313 XMM3) have only two observations each in UVW1.  As can be seen in Table~\ref{table4}, the two magnitudes observed are the same to within the errors for all three sources.  These objects can therefore be variable only to within this error based on the limited observations available here.  The fourth source however, Holmberg~II XMM1, is the only source with more than two observations in UVW1.   In this case, the magnitude appears to vary by 0.25 mag. over 8 years.  A $\chi^2$ test suggests that this source may indeed be variable ($\chi^2$ = 102 for 3 degrees of freedom, assuming a constant model).}

\subsection{OM Colors}

We performed a color comparison 
of the 100~pc regions with standard stars. 
Six color-color plots are presented in Figures~\ref{kurucz1}-\ref{kurucz3}. 
To create the stellar models, we used the Kurucz star library\footnote{http://www.stsci.edu/ftp/cdbs/grid/k93models} to estimate OM colors for stars with solar metallicity.
The fluxes were folded through the latest OM response matrices.
\citet{kun08} generated an OM Catalog and found that in general the OM colors 
match the Kurucz models well, except for late type stars, where a systematic offset is seen.  We note that not all objects were observed in every filter.  Therefore some ULXs appear in some plots and not in others.  If a ULX appears in one plot and not another, this simply means that the ULX was not observed in one or more of the color filters needed for the second plot.    

Since the accretion disk can have an important contribution to the color emission, we show the 
theoretical estimates from a standard accretion disk.  The UV emission in this case is described by a power-law with energy index 1/3 for any BH mass.
Therefore the UV colors do not change with the BH mass.
They change however when we include self-irradiation of the accretion disk. 
We followed \citet{san93} to calculate theoretical spectra from irradiated disks for BH masses up to 10000 M$_{\odot}$, 
and then estimated UV colors. We assumed accretion at the Eddington limit.
The theoretical OM color tracks are plotted as thick orange lines in Figures~\ref{kurucz1}-\ref{kurucz3}.  The errors in the figures are large, especially for the 7 ULXs where the 100~pc flux is compatible with single stars (not including the upper limits in Table~\ref{table2}).  {\bf The color information was sufficient to determine spectral type for only 2 sources: NGC4395 XMM1 and M101 XMM2.  We find a spectral type of B3I and O3 - O5V respectively.  Both estimates are close to the UVW1 predictions, but do not completely match them.}  We note that in many cases, our fluxes are likely not from single stars, as most ULXs are located in crowded fields (see Section~3.1 above).  We also note that the self-irradiated disk produces colors that are similar to young stars.  It is possible that a significant part of the detected UV flux for some ULXs plotted in 
Figs.~\ref{kurucz1} through \ref{kurucz3} is emission from the accretion disk.  This is consistent with the accretion disk contamination we discussed in Section 3.1.3. 

We note that the irradiation of the donor by the X-ray emission further complicates the optical photometric estimates.  
\citet{cop05} found that both the absolute optical magnitudes and the colors of ULXs 
(accretion disk plus companion) vary significantly with the mass of the BH. 
For example, for an O5V star companion, when the BH mass is varied between 10 and 1000~M$_\odot$, 
the V-band changes by 1.0 magnitude and the B-V color by 0.2.

\section{STARBURST99 MODELING AND COLORS}

In the previous Section we found that some ULXs are apparently located inside SFRs.
In most cases, bright SFRs are also found in the vicinity of ULXs.
Such regions usually contain young clusters or OB associations, which are strong UV sources.   Indeed these environments seem necessary for a ULX to subsist.   For example, IMBHs can form inside dense young clusters \citep{por02,gur04}.  In addition, stellar-mass black holes can produce super-Eddington luminosities only
if the donor is a massive star \citep{pod03}.  For this reason we explored any possible correlations between the ages and masses of our star forming regions and the ULX.   

We estimated ages and masses for the nearby star-forming regions,
using the Starburst99 code \citep{lei99} version~5.1. We simulated an instantaneous burst model 
with a Salpeter Initial Mass Function (IMF) and having exponent $\alpha=$~2.35, mass boundaries at 1 and 100 M$_{\sun}$ {\bf and a total cluster mass of 10$^6$M$_{\sun}$.}

The ULXs in this sample are located in different galaxies having a range of star formation rates (see WMR). 
However, \citet{win07} used the absorption features in XMM-Newton X-ray spectra
to measure the abundances for the ULXs in the WMR objects. 
The results are consistent with solar abundances for the entire sample, despite the wide
range of galaxy host properties.  
On the other hand, some authors found lower abundances for some ULXs \citep[e.g.][]{brass05,map09, ber10b, pin12}.       
Fortunately the models for young UV-bright stars are not sensitive to metalicity.  We therefore use solar abundances and do not expect large errors from this assumption.
Moreover, given the large error bars for the measured colors, 
we found that models with different IMF slopes or cutoff masses would not change our results.

\subsection{Starburst99 Modeling}

Apart from the 100~pc regions, we identified several bright SFRs 
close to most of our ULXs. The total number of regions for all 27 ULXs is 90.
We used the generated Starburst99 models to predict ages and masses for both the 100~pc and the star forming regions.  

In Figures~\ref{starburst1}$-$\ref{starburst3} we compare the OM colors of the SFRs with the Starburst99 models evolved to 900~Myr.
The Starburst99 model tracks are also plotted for predicted extinctions of A$_V=$ 0, 1 and 2.  Because most objects were not observed in every filter, the ULXs and associated SFRs are not the same from figure to figure.   With the exception of Holmberg~II and NGC~5408, all the galaxies 
were observed with at least three different filters.  We chose the colors that make up each axis in such a way to ensure
all regions are plotted in at least one plot. 
The measured colors were corrected for aperture effects as explained in Section~2.
We show the magnitude and direction of the maximum aperture correction applied to each plot with an arrow.

We estimated approximate ages by comparing the colors to the Starburst99 tracks.
When colors for a region were present in more than one plot, 
we used all the available data to make a best estimate for the age.  Three of the six color-color plots are degenerate with respect to the cluster age and extinction 
(Fig.~\ref{starburst2}, both panels and Fig.~\ref{starburst3}, right panel).
Therefore, when one color-color plot was degenerate, we used information from the other plots to establish a cluster age.  We found that the estimated ages agree reasonably well for different colors, to within the errors.

The mass of the cluster can be estimated from the age and extinction predicted from Figures~\ref{starburst1}$-$\ref{starburst3}.  To do this we first modeled the relationship between cluster age and UVW1 absolute magnitude for a 10$^6$~M$_{\sun}$ cluster having three extinction values (A$_V=$ 0, 1 and 2).  This model is shown in Figure~\ref{uvw1}.  We then find the UVW1 magnitude in this plot that corresponding to a cluster age and A$_V$ from the Figures~\ref{starburst1}$-$\ref{starburst3} in the analysis described above.  The magnitude from this plot will correspond to a mass of 10$^6$~M$_{\sun}$.  We finally scaled this mass based on our {\it observed} UVW1 magnitudes.  For example, a 50 Myr cluster with A$_V$ = 0 will have a UVW1 magnitude of -14 (See Figure~\ref{uvw1}).   If our {\it observed} UVW1 value is a factor of 10 fainter then we can predict that the mass is smaller than 10$^6$~M$_{\sun}$ by a factor of 10.   For the clusters with no reliable age estimate, we calculated an approximate mass, assuming a fixed cluster age of 50~Myr.

\subsection{Results from the Color Comparison}

The  age and mass estimates from the color-color plots are listed in Table~\ref{table3}. 
{\bf Columns~3 and 4 show the average extinction and age estimates of all surrounding star forming regions.}  Only extinction values from the non-degenerate color-color plots were used. 

In Column~2 of Table~\ref{table3} we show extinction values calculated using the column densities obtained by WMR from spectral fitting.  We assume E(B$-$V)~$=$~1 for a column of 5.5$\times$10$^{21}$~cm$^{-2}$.
By comparing the estimates from the X-rays with the OM color estimates in columns 3, 4 and 8,
it is evident that in general the X-ray estimates are much higher. 
The average absorption column density based on X-ray spectral analysis in WMR is 4.5$\times$10$^{21}$~cm$^{-2}$.
This corresponds to an extinction of A$_V=$~2.5, while most of our estimates are $\leq$1.
{\bf Our results suggest that the absorption is localized, in the immediate vicinity of the ULX (on scales less than the binary separation).}
\citet{kro04} proposed that some ULXs could be IMBHs inside dense molecular clouds and are fueled through Bondi accretion. 
For the most part, our color-color plots do not show strong extinction usually associated with dense interstellar media,
and therefore do not support this scenario.

In Columns~6-9 we list the results for the 100~pc regions.
In many cases, reliable ages could not be estimated because of low fluxes and large errors.
However, the ones that could be measured indicate relatively young ages.
Only NGC~253 XMM1 showed an age of $>$100~Myr.

Columns~10-13 of Table~\ref{table3} show the results for the closest SFRs to each ULX.
Thirteen (approximately half) are actually located inside these star forming regions, meaning that the star forming region overlaps with the 100~pc ULX regions.  This is true  
even if the intrinsic offset is quite large (e.g. $>$200~pc for NGC~253 XMM6). These 13 objects are marked in bold in column~10.
Four of the ULXs have age estimates $<$10~Myr.
Of these, the SFR located near NGC~4631 XMM1 and IC~342 XMM3 have relatively large masses ($>$10$^5$~M$_{\sun}$).
Apart from the ULXs located inside the SFRs, nine other ULXs have SFRs within 200~pc (21 in total out of 27).

The results for the youngest SFRs are shown in Columns~14-17.
Seventeen out of 21 ULXs 
have nearby regions younger than 10~Myr.
Only NGC~253 XMM1 does not have a nearby SFR younger than 50~Myr.

Finally, we consider the most massive SFRs found nearby the ULXs in our sample (cols.~18-21).
There are 16 SFRs with masses $>$10$^5$~M$_{\sun}$, but only 12 with age estimates.  Of these 12, roughly half (7) are younger than 50~Myr, including the central cluster in IC~342.  

The final row of Table~\ref{table3} shows the median extinctions, ages, offsets and masses for the five categories described above:  the average region, the 100~pc region, the closest, youngest, and most massive star forming regions.  We find that in general the closest star forming regions have stellar ages that are approximately 9~Myr and are offset from the ULX at an average distance of 87~pc.  On the other hand, we find that the majority of the most massive regions are 12~Myr in age and sit 240~pc from the ULX.  The youngest star formation regions tend to be 309~pc away and be approximatly 5~Myr, with the exception of NGC 253 X-1 which has a relatively old population of stars.

\subsection{Previously Known Star-forming Regions}

Some of the SFRs located nearby the ULXs in our sample were previously studied in the literature.  These are discussed in detail below.  

{\bf Holmberg~IX XMM1:} Holmberg~IX XMM1 (known as X-1) is associated with a bubble-like shell, 
and is located at the edge of a stellar cluster with estimated mass of 10$^3$~M$_{\sun}$ 
and older than 20~Myr \citep{gris06}. \citet{gris11} used {\it Hubble} data to estimate a color excess of E(B-V)$=$ 0.26. 
They found that the ULX is located inside an OB association with mostly young stars 
($<$20~Myr) and having mass $\sim$2$\times$10$^3$~M$_{\sun}$.
We could not make a reliable age estimate based on the OM colors due to insufficient data,
but our mass estimate (6$\times$10$^3$~M$_{\sun}$) is consistent with this result, 
given the older age we used (50~Myr).

{\bf NGC~1313 XMM3:} NGC~1313 XMM3 is located at the edge of a relatively young cluster, which we estimated to be $\sim$5~Myr old (for extinction A$_V\approx$~0.5, col.~15 in Table~\ref{table3}). 
\citet{liu07} found a similar result: age $\sim$10~Myr for a slightly larger extinction (A$_V\approx$~1).
\citet{tao12a} used Hubble data to estimate the age of the stellar association located 
very close to the ULX. Using a color excess of E(B-V)$=$0.18 they found a mixed population
of mostly young stars, in agreement with our estimate.

{\bf IC~342 XMM3:} For the central cluster where IC~342 XMM3 is located, we estimate an age of 5~Myr, 
and an extinction A$_{V}=$~2.5. This age is slightly younger than the one obtained by \citet{bok99},
and our estimated mass for the cluster, 10$^7$~M$_{\sun}$,  is almost double.
However, the distance to IC~342 used by these authors was smaller than our distance (3.9~Mpc), and this could account for the discrepancy.  For consistency our distances are taken from WMR.  

{\bf NGC~5408 XMM1} \citet{gris12} analyzed {\it Hubble} data for the OB association located close to NGC~5408 XMM1.
For an extinction A$_{V}=$~0.25 they estimated an age of 5~Myr and suggest that the ULX
could be a runnaway binary with a velocity of $\sim$~25$-$50 km~s$^{-1}$.
Unfortunately we only have images in two filters for this ULX, and could not get estimates of age for this object using OM.

\section{RUNAWAY SCENARIOS}

One of the main findings of this study is that many of the ULXs in our sample are located {\it inside} star forming regions rather than outside of them.  We note, however, that the OM has poorer resolution than the instruments used in many of these studies.  The rest of the ULXs (12 total), are indeed located nearby such regions.  In order to explain these rogue ULXs we explore the scenario where ULXs form in young stellar environments and are later ejected.  

\subsection{Stellar-mass X-ray Binary Runaway}

If the stellar mass black holes have super-Eddington accretion as described by \citet{beg02}, then \citet{pod03} shows that the observed X-ray luminosities can be obtained in systems 
with massive donors ($\gtrsim$~10~M$_{\sun}$, or stars earlier than $\sim$B2).
For this analysis we assume that the ULXs are stellar-mass X-ray binaries.  

ULX stellar-mass BH candidates can be found outside of the parent cluster if the birth of the black hole creates kicks of sufficient strength.   
\citet{sep05} estimated that $\sim$70\% of the X-ray binaries 
can be ejected from the cluster in this way. They also find that the younger the region ($\sim$10~Myr),  the more X-ray binaries are present outside the cluster.
For such clusters, typical offsets
are 10~pc$-$100~pc. Larger offsets (of $\sim$1000~pc) are found when the cluster is older ($>$50~Myr),
but the number of binaries is also much reduced in that case.  Recently, \citet{rep12} found that natal kicks similar to neutron stars can explain 
the distribution of black hole binaries in our Galaxy.  {\bf However, \citet{reid2011} and \citet{miller2009} use radio interferometry to measure the parallax of Galactic black holes and find that relatively small kicks are required to produce the velocities measured. } 
\citet{zuo10} simulated natal kicks for X-ray binaries using different models of clusters and a range of parameters.
They found that the offsets are in general consistent with the observational data from \citet{kaa04}.

\subsubsection{Estimating kick velocities}
To estimate the kick velocity, we followed the procedures of \citet{fry01}, who used  Galactic neutron star velocities to scale to black holes velocities.  {\bf We note that \citet{fry01} assume that black holes are formed after a neutron star is formed (i.e. not through direct collapse), which may not be the case if these ULXs are more massive black holes.} 

The neutron star velocities we used when we performed this calculation are an average of three more recent measurements by \citet{cor98}, \citet{arz02} and \citet{hob05}.  
\citet{cor98} found a velocity distribution
consisting of a two-component Maxwellian distribution with characteristic velocities of 175~km~s$^{-1}$ and 700~km~s$^{-1}$,
and a weight predominantly on the lower velocity component. The mean two-dimensional 
and three-dimensional velocities for this model are 331~km~s$^{-1}$ and 421~km~s$^{-1}$ respectively.
More recently, \citet{arz02} found a similar model but with characteristic velocities of 
90~km~s$^{-1}$ and 500~km~s$^{-1}$ and roughly equal weights.
This model implies more pulsars with higher velocities than the previous model.
The most recent study of \citet{hob05} fit the pulsar velocity distribution to a single Maxwellian distribution
with a mean three-dimensional velocity of 400~km~s$^{-1}$ and  no evidence for a multiple-component velocity.  The mean pulsar velocity we use for our calculations is 360~km~s$^{-1}$.

Using the \citet{fry01} method, we calculate kick velocities as:

$v=500($M$_{BH} + $M$_{comp})^{-1}$~km~s$^{-1}$.

The velocities are scaled to the total mass of the binary BH (M$_{BH}$) plus the companion (M$_{comp}$).
These velocities are more than twice the values found by \citet{zefa02} who use a similar method for the Antennae ULXs.  The discrepancy likely results because \citet{zefa02} use the values of \citet{cor98} for the initial neutron star velocity, while, we use these values averaged with those of \citet{arz02} and \citet{hob05}. Our calculation suggests that binary BH/companion star systems could have kick velocities that are quite large.  Indeed \citet{arz02} estimate that 50\% of the pulsars have velocities greater than 500~km~s$^{-1}$.

We note that the velocities we can measure for our ULXs (see below) are two-dimensional.  This means configurations of ULXs that have been kicked from the cluster toward or away from the viewer will not be taken into consideration in this estimate.  As a result the average velocity over all configurations of offsets 
is 4/$\pi$ times lower than the three-dimensional values $-$ at most a 27\% error on the velocity. On the other hand the neutron star velocities from which we base this calculation have a deviation well above this 27\% error. We therefore do not correct for the effect in our calculations.    

\subsubsection{Estimating X-ray binary lifetimes}

We estimate the lifetime of X-ray binaries using two mass transfer types \citep[e.g.][]{pat08}.
In the first case (A) we assume that the mass transfer starts early when the donor star is on the main sequence, 
while in the second (B), it begins after the donor has evolved off the main sequence.

It is much harder to observe a ULX in case B because the brighter/super-Eddington phase is much shorter \citep[by a factor in the range 10$-$100, see][]{pod03, rap05}.
On the other hand, the luminosities reached for case B are larger by an order of magnitude, thereby making the likelihood of seeing the ULX higher since the selection of possible companions is not restricted to massive stars.  The lifetime for the ULX in case B is equal to  
the lifetime of the donor star, as in \citet{zefa02}.

The lifetime calculation for case A requires a detailed stellar evolution simulation \citep[e.g.][]{pod03},
which is beyond the scope of this paper. A simple estimate of the {\it lower limit} can be obtained as follows.
We assume that the ULX luminosity is super-Eddington by a factor of 10 
and that the accretion efficiency, $\eta$, is approximately equal to 0.1. 
If the donor has an initial mass, M, and a luminosity, L (ten times the Eddington limit),
we estimate the lifetime for this system by taking the harmonic mean 
between the lifetime of the donor on the main sequence and the time T~$=\eta$Mc$^2$/L,
in which the entire mass of the stellar companion is accreted.
These two mass transfer cases are taken as limiting cases. 
The transfer can theoretically start at arbitrary times on the main sequence.
However, \citet{pou12} point out that there is a minimum delay between the time the binary is ejected
and the age of the donor, because the primary evolves inside the cluster before the supernova explosion.  The binary must therefore be at least 3~Myr old when the system is ejected. 
We take this into account when we calculate the offsets by subtracting 3~Myr from the donor's remaining lifetime.

\subsubsection{Estimating theoretical offsets}

Using the velocities and lifetimes calculated in this way we can estimate 
the {\it maximum} offsets from the star-forming regions for different BH and donor masses,
and compare them with the measured offsets for our sample. These calculations are bound by  two constraints:  the age of the companion and the accretion rate of the system. 
The ages of the companion stars are constrained from the Starburst99 models and observed OM data of the cluster from which they are thought to be born.
This in turn constrains the offset distances for the ULXs, assuming the stellar companion was acquired from the closest cluster.
The accretion rates are estimated using the bolometric luminosities from WMR.

1. {\bf Age estimates for the clusters:} Based on the derived age (presented in Table~\ref{table3}), 
we plot in the left panel of Figure~\ref{runaway1}
the largest mass possible for the companion, assuming it was formed in this cluster.
All objects with age estimates are plotted here.
When several star-forming regions were identified,
we used the data for the closest to the ULX, i.e. those presented in cols.~10-13 of Table~\ref{table3}.
For the case A transfer, we estimate offsets
based on the lifetime of the binary and assume super-Eddington luminosities for the accretion rate.  For case B the lifetime is the natural lifetime of the star.  In case B the accretion rate is negligible because it only comes into play at the final phase of the campanion star lifetime (consists of $\sim$100,000 years).
Only 9 ULXs could have companions with
masses $\gtrsim$~10~M$_{\sun}$ (earlier than type B1), capable of sustaining super-Eddington luminosities
on long timescales (case A mass transfer).

In Figure~\ref{runaway1} we compare the measured (projected) offsets with the theoretical offsets derived from the upper-limit calculation described in this section for different BH masses. {\bf We note that the theoretical lines presented in this plot do not include the potential of the cluster.  We estimate that the cluster contribution to the total velocity will be very small for large kick velocities ($>$ 40~km s$^{-1}$).  However for small kick velocities this contribution could be quite large and if the kick velocity is small enough the BH will not escape the cluster at all. } In the case A transfer the offsets plotted are consistent with BH masses $<$20~M$_{\sun}$.  For the case B transfer, larger offsets are predicted by the theoretical models, and most are consistent with BH masses $<$40~M$_{\sun}$, which we take as the mass limit for a BH formed from a supernova.
We also show the kick velocities for each donor mass at the bottom of the plot (in km~s$^{-1}$).

As an example consider NGC~247 XMM-1. For the closest star cluster region to this ULX (64~pc offset from the ULX position) we find a 12.5~M$_{\sun}$ donor that is ~10~Myr old (see the results for this object in Table~\ref{table3}).  This places the ULX firmly between the green and red dotted lines in Figure~\ref{runaway1}.  The lines represent the expected relationship for 20~M$_{\sun}$ and 10~M$_{\sun}$ black hole systems respectively based on our models.  This implies that for Case A mass transfer (dotted lines), the accretor has a mass of $\lesssim$15~M$_{\sun}$.  The ULX donor mass vs. offset is well below any of the case B mass transfer (solid lines) models, and therefore we have no constraint on the mass for this case.

In general our measurements follow the relationship between the mass of the companion and the measured offset distance
as suggested by the theoretical offset tracks for the case B (continuous lines).
We note that the errors in estimating the masses of the objects used for Figure~\ref{runaway1} are very large.  As mentioned in Section~2, the resolution of the OM is not sufficient to resolve the ULX and separate the ULX emission from the surrounding stars. Therefore the colors used to estimate the stellar mass are likely contaminated.  Moreover these plots only show the closest star forming regions to the ULX, when in fact, the ULXs could be related to other regions farther away.

2. {\bf Using the bolometric luminosities from WMR (the right panel of Figure~\ref{runaway1}):} 
This calculation does not require an estimate for the cluster or companion star age.  The offset calculated is based solely on the timescale of the accretion for the system.  The  
accretion rate is estimated from the bolometric luminosities of WMR.   Case B is not relevant here because the accretion time is such a small part of the lifetime of the star (100~Kys vs. 10~Myr).  
We compare the empirical offsets of the closest regions for all ULXs with those calculated (theoretical) using the bolometric luminosity method described here.  The theoretical tracks were calculated for different donor masses, 
and two different accretion rates:  one at the Eddington luminosity (dashed lines) and one at ten times this limit (continuous).
The plot is similar to Fig.~3 in \citet{kaa04}, but differs because these authors 
assumed the same kick velocity (10~km~s$^{-1}$) for any BH mass.
For low-velocity kicks, the ULX might not be able to escape if it formed in a dense cluster.
We mark the location of a kick velocity of 10~km~s$^{-1}$ in the plot by thick black lines
crossing the theoretical tracks. ULXs located on the right side of these tracks 
have a low probability to escape such a cluster.

In general, the ULXs follow the theoretical predictions, showing smaller offsets for larger luminosities. 
The super-Eddington tracks (continuous lines), as upper limits, permit larger offsets for the same luminosity. This is because the kick velocity is higher for smaller black holes (see the top X-axis for the offset relationship to BH mass).  A comparison of the donor mass plot with the bolometric luminosity plot suggests that the latter calculation predicts smaller offsets.  This might be because the donor mass plots include models for Case B, where the luminosity models do not.  Most of the ULXs plotted are consistent with the super-Eddington case,
but not with accretion at the Eddington limit. There are a few ULXs with high luminosities 
that show larger offsets than permitted even for the super-Eddington accretion.

NGC~247 XMM1 has bolometric luminosity of 7.1$\times$10$^{39}$~erg~s$^{-1}$.
For sub-Eddington accretion, the corresponding mass of the BH is $>$55~M$_{\sun}$,
and the offset is too large for any companion mass.
If the accretion is super-Eddington, the companion could be a massive star ($\sim$20~M$_{\sun}$).

\subsection{IMBH Runaway}

If ULXs contain IMBHs then they cannot be ejected from dense clusters by natal kicks alone, 
because the kick velocity is smaller than the required ejection velocity.
\citet{millcol04} presented other mechanisms that can, in principle, deliver higher kicks.
The authors discuss the scenario of three-body interactions in young clusters,
and find that significant kicks ($>$10~km~s$^{-1}$) can be achieved even for a large BH (1000~M$_{\sun}$),
if the interloper is a young massive star (an O star).
The kick velocity will depend also on the binary orbital velocity, which can be estimated assuming Roche lobe overflow.

\citet{pou12} also suggested three-body interactions to explain the association between the Antennae ULXs and very young clusters.
The authors argue that the natal kicks cannot explain the data because the binary cannot be ejected before at least 3 Myr
(the life of a massive star). This leaves no time for the binary to reach the present position outside the cluster.
\citet{map11} performed simulations of three-body interactions in massive clusters.
They investigated ejections of black hole binaries with large masses,
capable of achieving the high accretion rates required to explain ULXs (up to 80~M$_{\sun}$).
They found that almost half of these binaries are expelled within 10~Myr,
and that the offsets are consistent with ULX observational data.

Following \citet{millcol04}, we calculate theoretical offsets using velocities estimated for a three-body interaction kick.
We estimate the X-ray binary lifetime using calculations from the previous section. 
As before, we compare the theoretical offsets with the measured offsets for our ULXs, 
for the two different constraints (donor age and ULX bolometric luminosity).
The results are presented in Fig.~\ref{runaway2}. For this calculation the accretion rate is assumed to be sub-Eddington.
The cluster age estimated using Starburst99 modeling and OM observations constrains the age of the binary companion,
but for the interloper we assume a fixed mass of 20~M$_{\sun}$.  A young O-type star is needed to create kicks large enough to eject the IMBH and, further, kick the IMBH to the offsets observed.  In addition the selection of a young interloper with high mass is in line
with our goal to estimate {\it upper limits} for the offsets.
We note that the interloper does not have to be a massive star, it could also be a 20~M$_{\sun}$ BH.

We consider again Case A and Case B mass transfer. Here too, Case A assumes that the mass transfer starts early when the donor star is on the main sequence and Case B assumes it begins after the donor has evolved off the main sequence.  For each mass transfer case, we also look at two accretion rates:  one at the Eddington limit and one at 10\% of the Eddington limit.  As we did before for natal kicks, we present the results used for the donor star calculation and bolometric luminosity calculation in Fig.~\ref{runaway2}.

1. {\bf Using the age estimates for the clusters (left panel of Fig.~\ref{runaway2}):} For both case A and case B mass transfer 
significant offsets ($>$10~pc) are possible even for BH masses of a few 1000~M$_{\sun}$, 
and most of our data are consistent with this scenario.  On the other hand,  the velocities predicted for such BHs are
smaller than 10~km~s$^{-1}$ (they are shown on the green line at the bottom of the plot), implying that the system can not realistically escape the cluster. 
For NGC~247 XMM1 the offset is consistent with a BH mass $<$700~M$_{\sun}$ for case A 
and $<$2000~M$_{\sun}$ for case B (if it somehow escapes the cluster).  

2. {\bf Using the X-ray-based bolometric luminosities from WMR (right panel of 
Figure~\ref{runaway2}):} 
As with the stellar-mass BH scenario, only case A is considered for the bolometric luminosity offset calculation.
Just like Figure~\ref{runaway1}, the constraint based on the luminosity suggests lower offsets 
than the one based on the donor mass.
Very few of the ULXs located on the left side of the plot have offsets 
consistent with the bolometric luminosity predictions. Those ULXs that do show some consistency, only correspond to models of accretion close to the Eddington limit (dashed lines).
Moreover, the BH masses for these ULXs do not exceed 80~M$_{\sun}$.
Again we mark the location of a kick velocity of 10~km~s$^{-1}$ in the plot by thick black lines.
For accretion at 10\% of the Eddington limit (continuous lines) the offsets and ejection velocities
are too small to explain the data.
The offset and luminosity for NGC~247 XMM1 are not consistent with a ULX ejected 64~pc from the cluster via three-body interactions.

\section{ DISCUSSION}

\subsection{ULXs Located Inside Star-forming Regions (Clusters)}

Simulations have shown that in young and dense clusters, IMBHs can form through core collapse, the black holes having masses of 0.1\%$-$0.2\% of the cluster mass \citep{por02,gur04}.
We assume therefore that IMBHs can form only in very dense clusters with masses $>$10$^5$~M$_{\sun}$.
One such example was thought to be M82~X-1. \citet{por04} found an X-ray position coincident 
with the cluster MCG~11, the cluster having  a mass of 10$^{5.5}$~M$_{\sun}$, and a half-light radius of 1.2~pc.
Recently, however, \citet{voss11} showed that the ULX is actually offset by 16~pc from the super-cluster
(assuming a distance of 5.2~Mpc for M82).  Supporting the dense cluster theory further, the authors also discovered a new ULX in NGC~7479,
and showed that it is a true example of a IMBH located inside a super-cluster with mass 
in excess of 7$\times$10$^{5}$~M$_{\sun}$ and half-light radius of $\sim$20~pc.
Two additional ULXs in the Antennae galaxy were also shown to be near young super-clusters \citep{clark11}.

Once the IMBH is formed, it still needs to capture a companion star to become a ULX.
\citet{ble06} studied the probability of an IMBH becoming a ULX using simulations 
of young ($<$100~Myr) clusters. For dense clusters they found that the probability 
of capture through binary exchange is relatively high, but that in general the resulting binary would appear as a ULX
for only 2\% of the total transfer time. This implies a typical ULX lifetime shorter than 0.1~Myr,
and a low probability to observe such a ULX. \citet{baum06} considered higher-mass IMBHs
and found higher probabilities. Their results show that the captured companions  
are massive main-sequence stars, capable of producing high X-ray luminosities over timescales of $\sim$10~Myr.

The results in Table~\ref{table2} suggest that 18 of the ULXs in our sample
are located inside OB associations or SFRs. 
There are 3 ULXs which apparently reside inside young SFRs according to the results in Table~\ref{table3} ($<$10~Myr, Columns 6$-$9).
However the cluster masses are not large enough to contain young super-clusters in which IMBHs might grow.
The estimates in Table~\ref{table3} show that when detection and age estimates
were possible, the majority of the stellar populations have ages $\lesssim$20~Myr.
This is true both for the 100~pc regions (col.~6) and the SFRs that contain ULXs (ULXs in SFRs are in bold in col.~10, corresponding ages are shown in col.~11).
In such young regions, the donors are likely massive stars.  In addition, super-Eddington luminosities
can theoretically be achieved with stellar-mass BHs rather than IMBHs, as suggested by \citet{pod03}.

IMBHs can also form in older globular clusters through the much slower mechanism of BH mergers \citep{mill02}.
If a larger BH ($\gtrsim$50~M$_{\sun}$) is initially formed in the cluster, it can grow through mergers with smaller black holes.
Hardening of binaries through three-body interactions cause the mergers.
The process is thought to be slow and may be inefficient, as many interlopers are ejected.
\citet{lear06} built onto this mechanism gravitational radiation and successive mergers of binary and single BHs \citep{gul04}, as well as  binary-binary interactions.
In this case, the host clusters for ULXs should be older clusters ($>$100~Myr), 
and the IMBH companions later type stars.
Only NGC~253 XMM1 seems to be located inside an older stellar population according to our estimates in Table~\ref{table3}.
Given the resolution of the OM instrument, we cannot tell if compact clusters are present.  Moreover, the UV is not very well suited to detect old stellar populations.

\subsection{Nearby Clusters and the Runaway Scenario}

If our ULXs are not apparently located inside super-clusters, 
they could still have formed in such clusters and later been ejected.
As mentioned before, it is difficult with OM to identify dense super-clusters inside star-forming regions. 
However, some of the star-forming regions located nearby (Table~\ref{table3}) 
are young and massive enough to host such clusters.
If ULXs are super-Eddington sources, they require massive stars as companions 
\citep[$\gtrsim$~10~M$_{\sun}$, type B2 or earlier, see][]{pod03}.
We have found that the majority of ULXs 
(17, see cols.~14-17 of Table \ref{table3}) are located nearby SFRs young enough to host such massive stars.
\citet{rap05} estimate that in star-forming regions, 
ULXs can have a typical life of 10 to 30~Myr after a starburst, but this lifetime can be as long as 100~Myr.

The runaway scenario is normally used as evidence for the stellar-mass BH in ULXs,
the argument being that more massive BHs are not easily ejected from clusters.
However, three-body interactions can eject even 1000~M$_{\sun}$ BHs.
In Section~5 we have therefore tested runaway scenarios for both stellar-mass BHs and IMBHs.
Specifically the two scenarios are: 1) stellar-mass black hole binaries emitting at ten times the Eddington limit
ejected by natal kicks, or 2) ULXs with IMBHs accreting within the Eddington limit,
ejected by three-body interactions in dense environments.
We have found that the first scenario fits the data much better (Fig.~\ref{runaway1}).
However that does not completely rule out some ULXs in this sample as examples of the latter. 
The left panel shows that the IMBH scenario only works for BH masses up to $\sim$700~M$_{\sun}$, 
while in the right panel the offsets cannot be explained by this scenario (assuming case A only).
\citet{map11} showed with simulations that massive BHs (up to 80~M$_{\sun}$) can be easily 
ejected via three-body encounters inside dense clusters, however IMBHs have a low probability of being ejected
because this requires a high-mass interloper.

Finally, another scenario in which ULXs containing IMBHs are associated with star-forming regions 
but not with dense clusters has been proposed by \citet{mill02}.
An IMBH formed in the core of a globular cluster might be released in the galactic disk when the cluster disperses.
If the IMBH encounters a molecular cloud, the interaction could precipitate the collapse of the molecular cloud and produce star formation.
In the newly formed star cluster the massive stars would sink towards the center and be captured by the IMBH.
This scenario could explain the fact that many ULXs are close to star-forming regions but not to dense stellar clusters.

\subsubsection{Problems with natal kicks}

There are a few problems with the natal kicks scenario.
Large kick velocities, as used in this paper, can easily disrupt the binary.
Even if it survives, the orbit will become highly eccentric.
\citet{rap05} and \citet{map11} note that such systems will have an unstable mass transfer 
if the mass of the donor is larger than the BH. Moreover, the mass-transfer phase is very short, making observing these objects in the ULX phase unlikely.

Our assumption that the accretion is super-Eddington is not very probable if the donor is a main sequence star 
with a low mass. \citet{pod03} have shown that even if we assume that 
the accretion is super-Eddington, luminosities $>$~10$^{39}$~erg~s$^{-1}$ are only obtained
for donors with masses $\gtrsim$~10~M$_{\sun}$, which supports this assumption.

The natal kick velocities for BHs depend on the amount of fallback associated with the supernova explosion.
For lower-mass BHs the velocities will be lowered with the amount of fallback \citep{bel04}.
Moreover, kicks are not expected for higher-mass BHs that form without an explosion.
\citet{bel04} estimated that the limit on the progenitor mass is $\sim$40~M$_{\sun}$.

\subsubsection{Newtonian and recoil kicks}
Newtonian kicks can eject lower-mass IMBHs from clusters that are not very dense 
(with escape velocities $\lesssim$~20~km~s$^{-1}$, as discussed in Section 5.2). However IMBHs can be subject to higher velocity, recoil kicks during BH mergers.
Recently, important advances in numerical relativity have allowed precise calculations
of the dynamics of BH mergers. These results show that the asymmetry 
during the plunge phase of the inspiral can produce ``recoil kicks'' with velocities as high as 4000~km~s$^{-1}$ 
\citep[e.g.][]{bak08}. The kick velocities depend on the mass ratio and spins of the black holes.  The velocities are largest when the BHs are of comparable mass and both have large spins. 
These large kicks have very important implications on IMBH growth in clusters.
\citet{holl07} showed that IMBHs with masses $<$~500~M$_{\sun}$ are very easily ejected, 
unless the merging companion BH has a mass $<$~10M$_{\sun}$.
\citet{bak08} obtained different results, however. They found that even a $\sim$~200~M$_{\sun}$ BH
can sustain radiative recoil kicks from a merger and never leave the cluster. This is the same limiting mass that Newtonian three-body kicks can eject \citep{lear06}.  {\bf \citet{jonker2010} was the first to propose this scenario for the ULX, CXO~J122418.6$+$144545.}

A slightly different scenario for some ULXs is a recoiling supermassive BH. 
\citet{fuj09} showed that accretion from the interstellar medium in the galactic disk can easily generate luminosities similar to ULXs.

\section{CONCLUSIONS}

\begin{figure}
\rotate
\epsscale{0.7}
\plotone{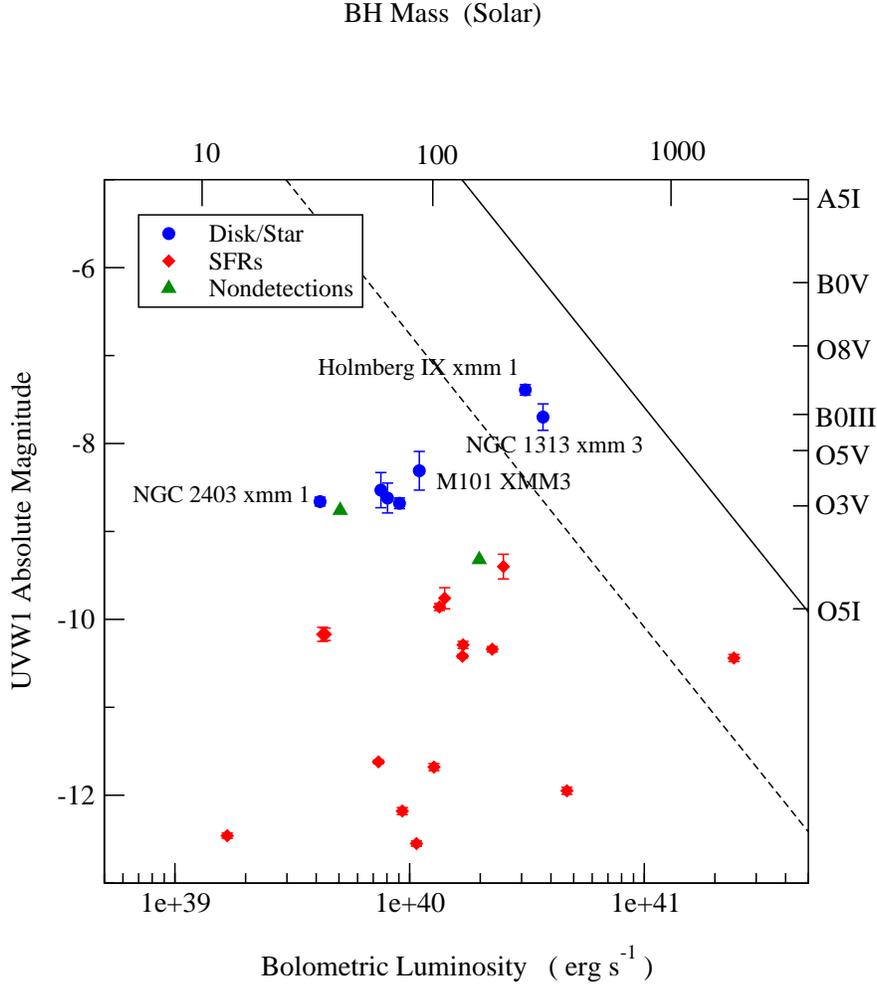}
\caption{
The UVW1 magnitudes measured in 100~pc regions centered on ULXs are plotted
against the bolometric luminosities estimated from the X-ray observations in WMR.
Possible single star detections as defined in Table~\ref{table2} 
are shown as circles while stellar associations are shown as diamonds.
Only four of these we estimate to have single stars as UV counterparts (see details in Section~3.1),
and these are labeled.
We also show with triangles the upper limits for ULXs with no detection (S/N~$<$~3).
Most UVW1 errors are smaller than the size of the symbols.
The straight line shows theoretical values for ULXs assuming the UVW1 fluxes 
are produced by a standard accretion disk accreting at the Eddington limit.  The dotted line shows the theoretical irradiated disk emission.
The marks on the top horizontal axis show BH masses corresponding to the Eddington
luminosities below. Similarly, on the right vertical axis we mark star types
corresponding to the UVW1 magnitudes on the left axis.
} \label{bh}
\end{figure}

\begin{figure}
\rotate
\epsscale{0.9}
\plottwo{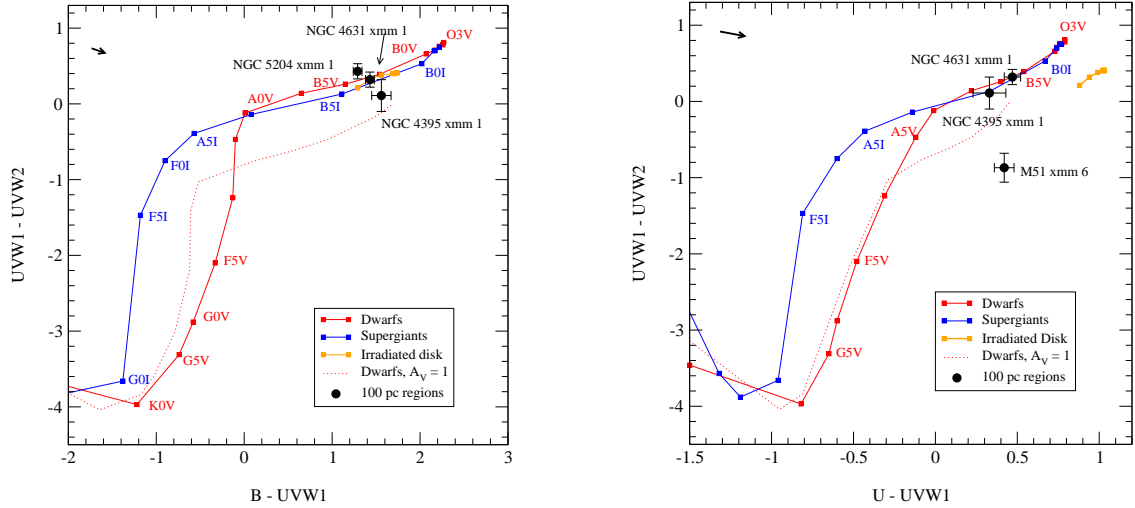}{f4b.eps}
\caption{
Color-color plots for the 100~pc regions in OM filters.
The plotted lines correspond to solar metallicity dwarfs and supergiants from the Kurucz library.
The dotted lines are reddened models (A$_V=$~1) for white dwarfs.
The short orange lines were calculated for ULXs with BH masses ranging from 
5 to 10000 M$_{\odot}$, assuming a standard accretion disk and self-irradiation (see text for details). The arrow at the top left corner marks the maximum magnitude and direction of the aperture correction.
} \label{kurucz1}
\end{figure}

\begin{figure}
\rotate
\epsscale{0.9}
\plottwo{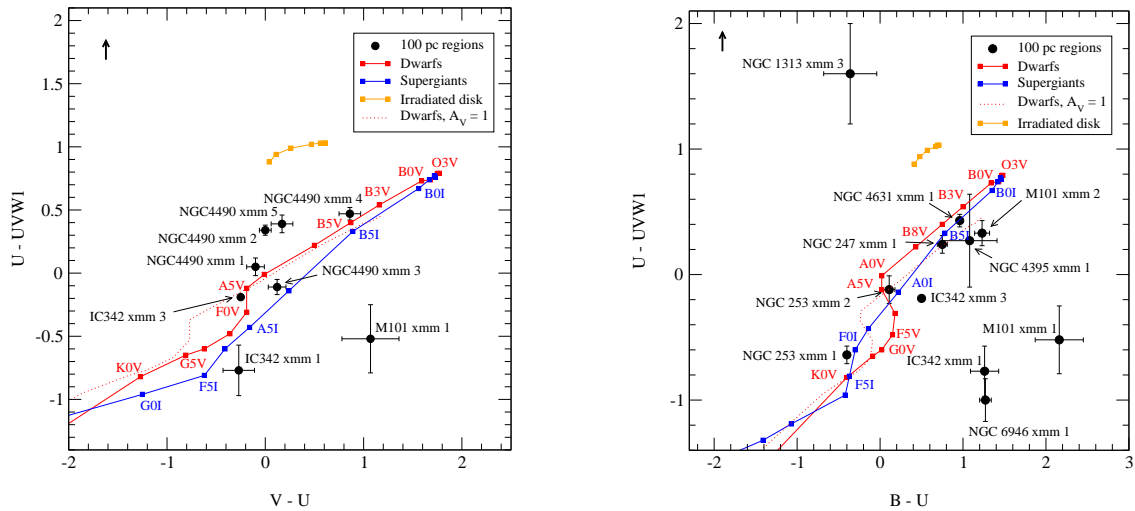}{f5b.eps}
\caption{
Color-color plots for the 100~pc regions, the symbols and tracks are the same as in Figure~\ref{kurucz1}.  The arrow at the top left corner marks the maximum magnitude and direction of the aperture correction
} \label{kurucz2}
\end{figure}

\begin{figure}
\rotate
\epsscale{0.9}
\plottwo{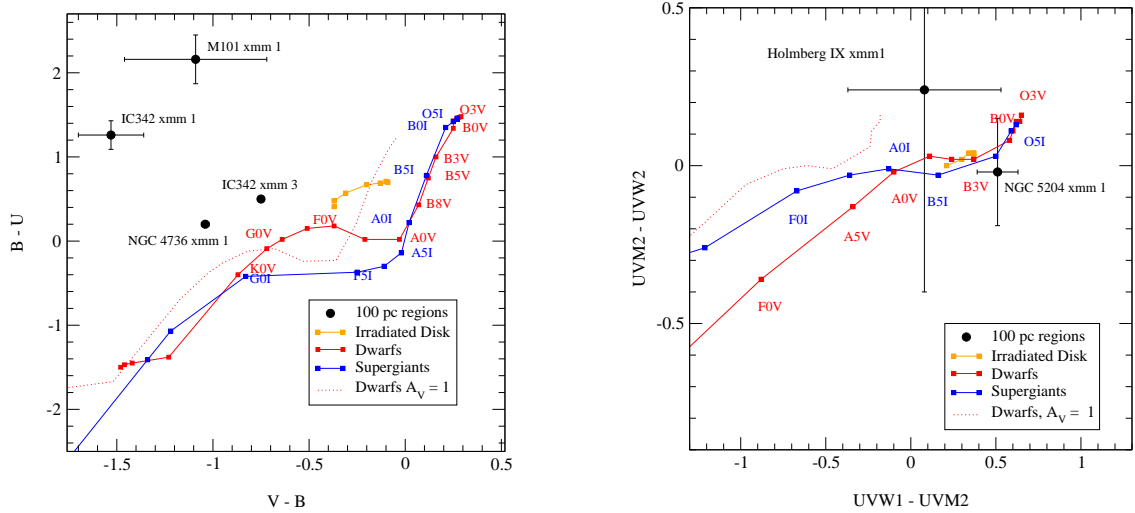}{f6b.eps}
\caption{
Color-color plots for the 100~pc regions, the symbols and tracks are the same as in Figure~\ref{kurucz1}.
} \label{kurucz3}
\end{figure}

\begin{figure}
\rotate
\epsscale{0.9}
\plottwo{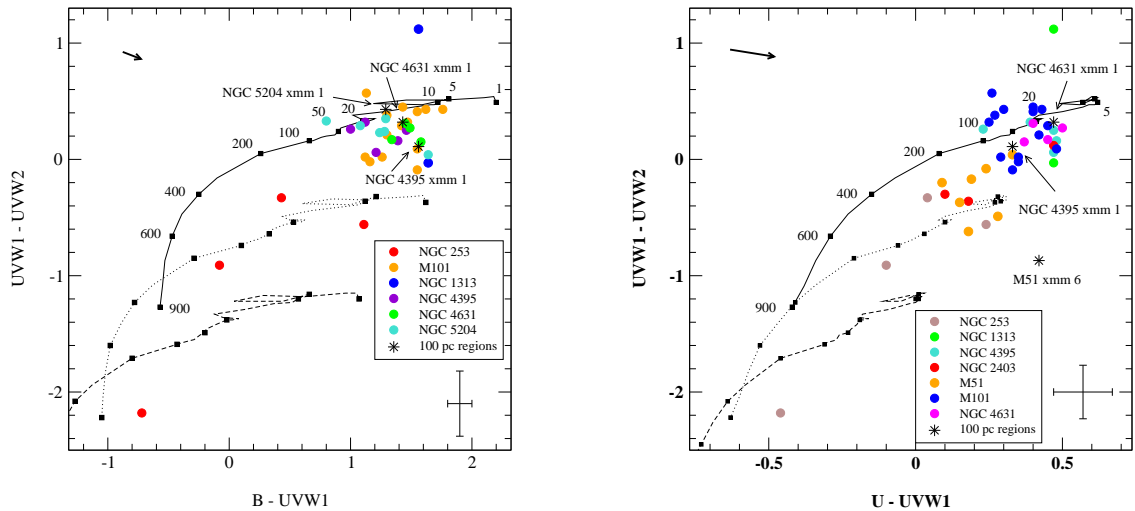}{f7b.eps}
\caption{
Color-color plots for the 100~pc regions and nearby SFRs observed with OM filters.
The plotted tracks are Starburst99 instantaneous models with solar metallicity.  The data points for each track represent ages ranging from 1 to 900~Myr.  
We also plot reddened tracks for A$_V=$~1 (dotted) and A$_V=$~2 (dashed), besides tracks with no extiction A$_V=$~0 (solid). The arrow at the top left corner marks the maximum magnitude and direction of the aperture correction. Finally we plot an average error bar over all data points in the right bottom corner of each plot.  
} \label{starburst1}
\end{figure}

\begin{figure}
\rotate
\epsscale{0.9}
\plottwo{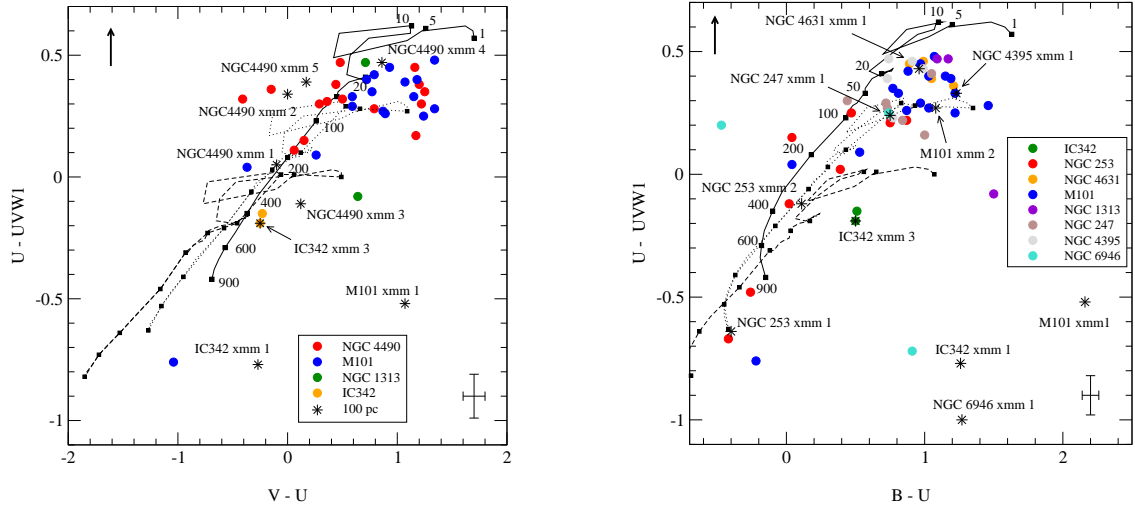}{f8b.eps}
\caption{
Color-color plots, the Starburst99 tracks are the same as in Figure~\ref{starburst1}.  We plot an average error bar over all data points in the right bottom corner of each plot. The arrow at the top left corner marks the maximum magnitude and direction of the aperture correction
} \label{starburst2}
\end{figure}

\begin{figure}
\rotate
\epsscale{0.9}
\plottwo{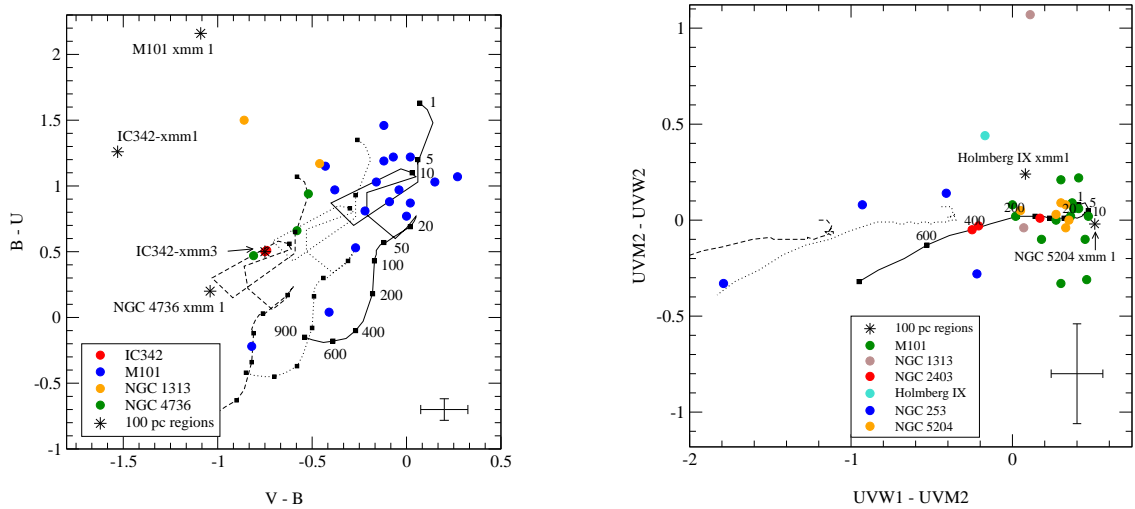}{f9b.eps}
\caption{
Color-color plots, the Starburst99 tracks are the same as in Figure~\ref{starburst1}. We plot an average error bar over all data points in the right bottom corner of each plot.
} \label{starburst3}
\end{figure}

\begin{figure}
\rotate
\epsscale{0.8}
\plotone{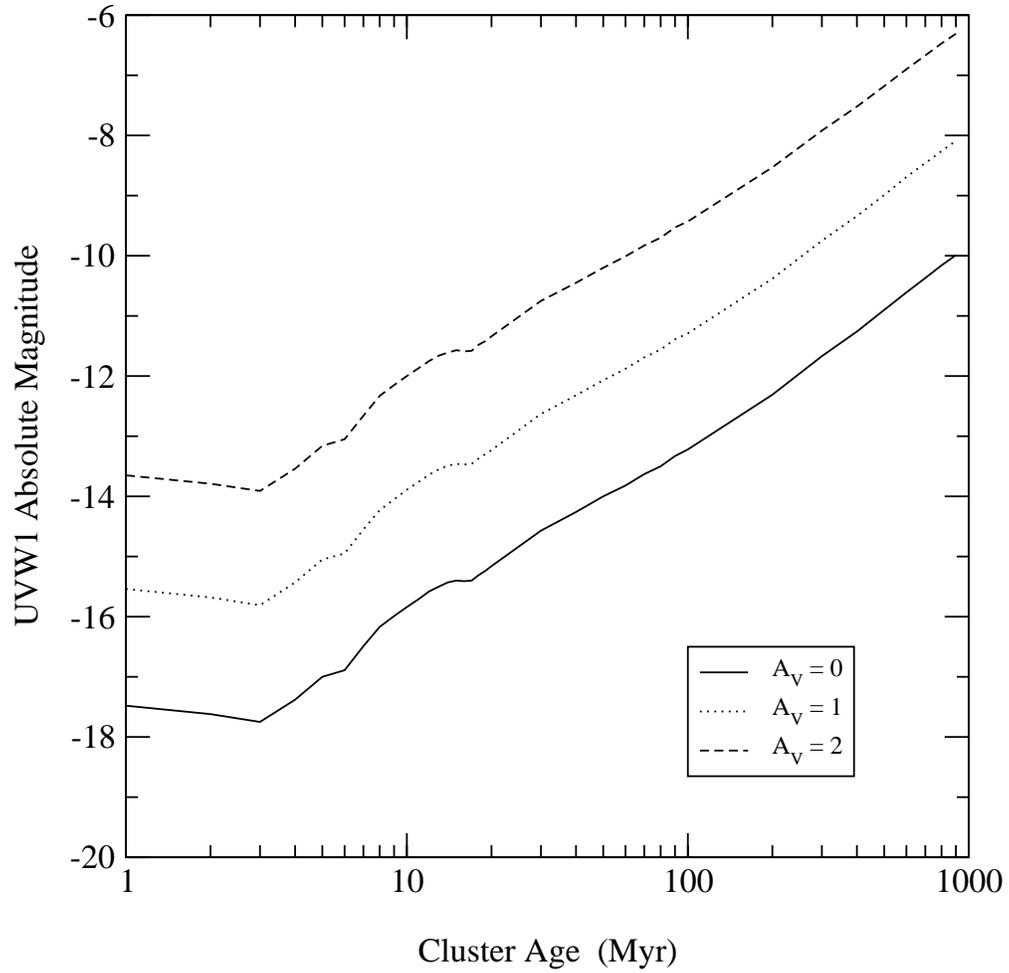}
\caption{
The UVW1 magnitude evolution of the Starburst99 model with a cluster mass of 10$^6$~M$_\odot$.
This was used to estimate masses for the clusters with reasonable age and extinction estimates, 
by scaling from the Starburst99 model.
} \label{uvw1}
\end{figure}

\begin{figure}
\rotate
\epsscale{1.0}
\plottwo{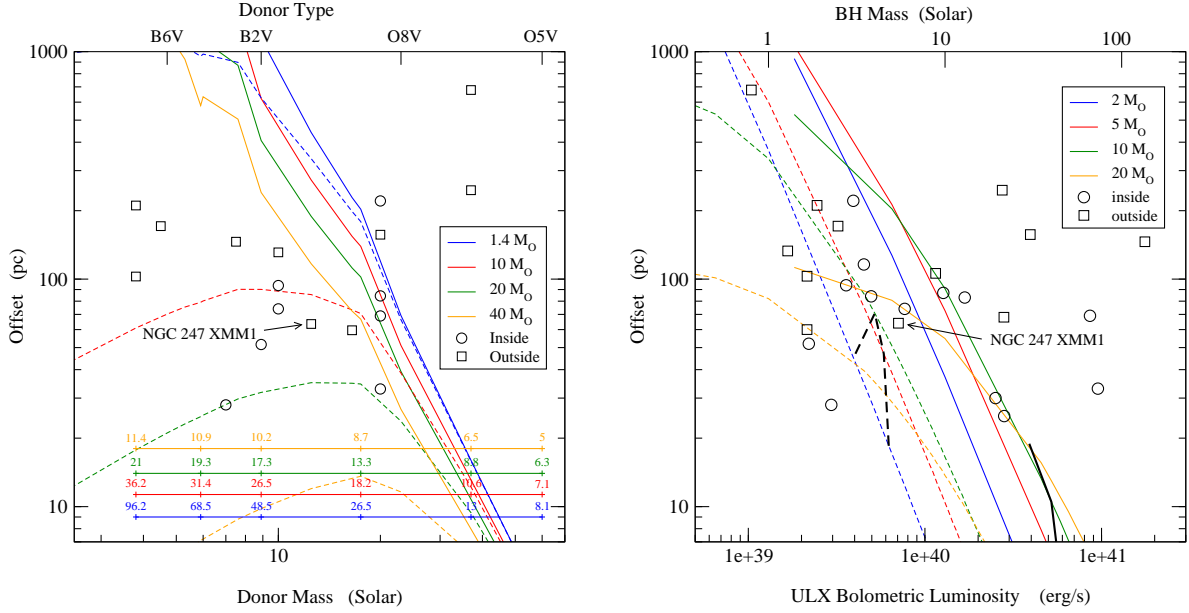}{f11b.eps}
\caption{
Diagnostic diagrams for the runaway binary scenario.
For the data points we used the distances and star type estimates for the closest SFR (Table~\ref{table3}, cols.~9 and 11).
{\it Left}: the tracks correspond to different BH masses. 
In some cases the ULXs appear to be located inside the SFRs (given the resolution of the OM).
These are plotted as circles.
The solid lines correspond to case B mass transfer, and the dashed lines for the case A.
The horizontal lines at the bottom of the plot show the kick velocities for each BH mass.
{\it Right}: Here the tracks correspond to different donor masses. 
On the x-axis we plot both luminosities (bottom) and the corresponding BH masses (top),
assuming accretion at 10 times the Eddington limit.
Here all the lines are for case A mass transfer. 
The solid lines correspond to super-Eddington accretion,
while the dashed lines for accretion at the Eddington limit.
We plot with squares the ULXs that are associated with massive clusters ($>$10$^5$~M$_\odot$).
The black thick tracks mark the limiting kick velocity of 10~km~s$^{-1}$,
for the Eddington limit (dashed) and for super-Edington accretion (continuous).
} \label{runaway1}
\end{figure}

\begin{figure}
\rotate
\epsscale{1.0}
\plottwo{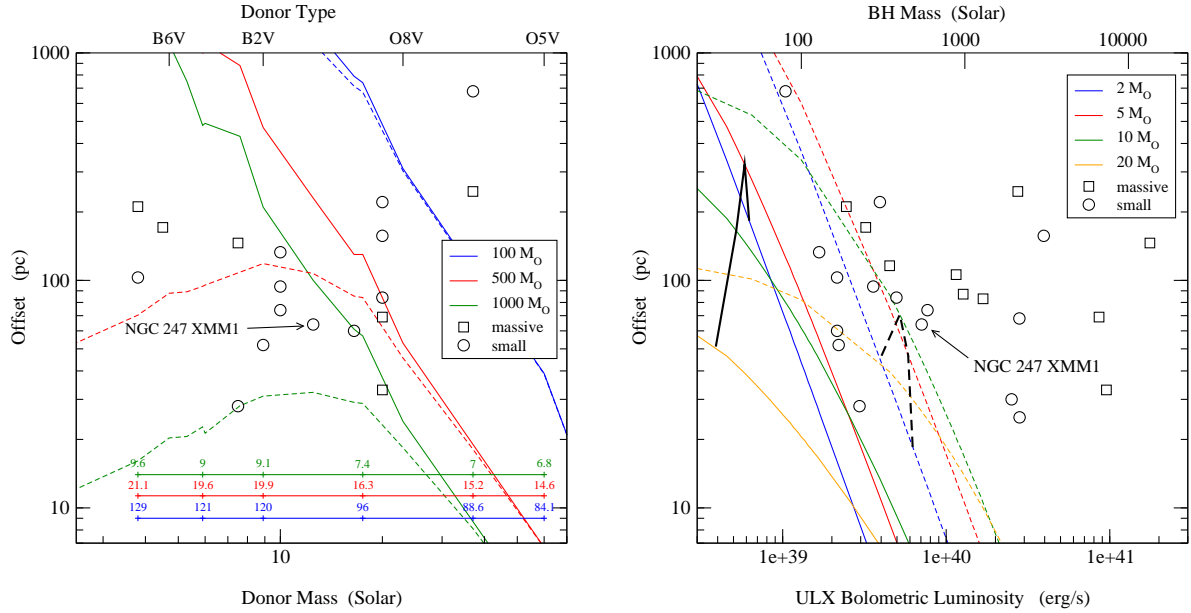}{f12b.eps}
\caption{
Diagnostic diagrams for the runaway IMBH scenario.
These plots are similar to Fig.~\ref{runaway1}, but now we assume that the ULXs contain IMBHs,
and the kicks are due to three-body interaction inside clusters.
As in Fig.~\ref{runaway1}, the ULXs that appear to be located inside SFRs are shown as circles.
{\it Left}: as in Fig.~\ref{runaway1}, the solid lines correspond to case B mass transfer, and the dashed lines correspond to case A,
but now we assume accretion at 0.1 the Eddington limit.
As in Fig.~\ref{runaway1}, we show ejection velocities on the horizontal lines at the bottom.
{\it Right}: the solid lines correspond to accretion at 0.1 the Eddington limit,
while the dashed lines for accretion at the Eddington limit.
The BH masses marked on the top x-axis correspond to the 0.1 Eddington accretion case.
As in Fig.~\ref{runaway1}, the black thick tracks mark the 10~km~s$^{-1}$ kick velocity limit
} \label{runaway2}
\end{figure}

The summary of the main results presented in this paper is as follows:

\begin{itemize}

{\bf \item Using UVW1 photometry, we have found that 7 sources with UVW1 magnitudes measured from 100~pc regions have emission that is consistent with a single star (see also Table~\ref{table2}, col.~11 and Figure~\ref{bh}).
Indeed for four of these objects our UVW1-predicted spectral types are identical to those found in the literature using other methods. We note that all of these objects (even those with consistent published data in the literature) may be contaminated by or dominated by an accretion disk component.  Indeed for at least two sources the UV emission likely comes predominantly from an accretion disk (see also bullet~2).  

\item By comparing the theoretical UV fluxes expected from accretion disks with the UVW1 measurements,
we found that, for most of the ULXs in our sample (esp. NGC~1313 XMM3 and  Holmberg~IX XMM1),
a significant part of the measured UV flux could come from accretion disk emission (Fig.~\ref{bh}).  However, even if the accretion disk is at its brightest (e.g. irradiated),  it is not bright enough in most cases to be responsible for all of the UV emission.  In those cases (the cases in Table~2, column~11 where the flux is consistent with multiple sources), it is likely that a star forming region, an accretion disk, and a donor star contaminates the aperture, none of which can be isolated using the UVW1 photometry alone. }

\item We looked at 3-5 SFRs around most ULXs.  We derived statistics for the closest SFRs to the ULX, the youngest SFRs around the ULX, and the most massive SFRs around the ULX.  Roughly half of the closest SFRs actually overlap with the ULX.  Of the youngest star forming regions, 17 out of 21 are less than 10~Myr in age, which might imply that young regions are intrinsically related to ULXs.  Finally there are also 17 sources with at least one region that is more massive than 10$^5$ M$_{\odot}$.  

\item OM color-color plots show relatively little reddening for most of the SFRs close to our ULXs, implying that they are likely not IMBHs accreting from molecular clouds.
The extinction is also, in general, much less than suggested by the absorption columns obtained from X-ray modeling, indicating that the X-ray absorption is located close to the ULX.

\item We tested runaway scenarios for both stellar-mass BHs and IMBHs.
Specifically the two scenarios are: 1) stellar-mass black hole binaries emitting at ten times the Eddington limit
ejected by natal kicks, or 2) ULXs with IMBHs accreting within the Eddington limit,
ejected by three-body interactions in dense environments.
We have found that the first scenario fits the data best (Fig.~\ref{runaway1}).  On the other hand, that does not completely rule out that some ULXs in this sample are IMBHs, though the masses we found in this analysis were somewhat small. 

\end{itemize}

\acknowledgments

We would like to thank R. Mushotzky and K. Kuntz for helpful discussions and suggestions. We are also appreciate the helpful comments from the anonymous referee, which have significantly improved the quality and clarity of this paper.
J. T. gratefully acknowledges support from the Hodson Trust Fellowship Program at St. John's College. 
This research has made use of the NASA/IPAC Extragalactic Database (NED) 
which is operated by the Jet Propulsion Laboratory, California Institute of Technology, under contract with the National Aeronautics and Space Administration.


\pagestyle{empty}  
\begin{deluxetable}{ccccrcc}
\tablecolumns{6}   
\tablewidth{0pt}                                                                                   
\tabletypesize{\tiny}  
\tablenum{1}                                                                                       
\tablecaption{OM observations\label{table1}}          
\tablehead{
\colhead{Gal} & \colhead{D} & \colhead{A$_V$} & \colhead{Obs ID} & \colhead{Date} & \colhead{Filters} \\
\colhead{} & \colhead{(Mpc)} & \colhead{} & \colhead{} & \colhead{(s)} & \colhead{} & \colhead{}\\
\colhead{(1)} & \colhead{(2)} & \colhead{(3)} & \colhead{(4)} & \colhead{(5)} & \colhead{(6)}\\
}
\startdata

NGC 247	&	3.09	&	0.06	&	0601010101	&	2009 Dec 27	&	U, B, UVW1, UVM2	\\
NGC 253	&	3.73	&	0.06	&	0304850901	&	2006 Jan 2	&	U	\\
$\cdots$	&	$\cdots$	&	$\cdots$	&	0110900101	&	2000 Dec 13	&	B, UVW1, UVM2	\\
$\cdots$	&	$\cdots$	&	$\cdots$	&	0152020101	&	2003 Jun 19	&	B, UVW1, UVW2	\\
NGC 1313	&	4.17	&	0.34	&	0106860101	&	2000 Oct 17	&	V, UVW1	\\
$\cdots$	&	$\cdots$	&	$\cdots$	&	0405090101	&	2006 Oct 15	&	B, UVW1, UVW2	\\
$\cdots$	&	$\cdots$	&	$\cdots$	&	0150280701	&	2003 Dec 27	&	U, UVM2	\\
IC 0342	&	3.90	&	1.73	&	0093640901	&	2001 Feb 11	&	UVW1	\\
$\cdots$	&	$\cdots$	&	$\cdots$	&	0206890101	&	2004 Feb 20	&	B, UVM2	\\
$\cdots$	&	$\cdots$	&	$\cdots$	&	0206890201	&	2004 Aug 17	&	U, UVW1, UVM2	\\
$\cdots$	&	$\cdots$	&	$\cdots$	&	0206890401	&	2005 Feb 10	&	V, UVW1, UVM2	\\
NGC 2403	&	3.56	&	0.12	&	0150651101	&	2003 Apr 30	&	UVW2	\\
$\cdots$	&	$\cdots$	&	$\cdots$	&	0150651201	&	2003 Sep 11	&	U	\\
$\cdots$	&	$\cdots$	&	$\cdots$	&	0164560901	&	2004 Sep 12	&	UVW1, UVM2	\\
Holmberg II	&	2.70	&	0.10	&	0112520701	&	2002 Apr 16	&	UVW1	\\
$\cdots$	&	$\cdots$	&	$\cdots$	&	0561580401	&	2010 Mar 26	&	UVW1	\\
Holmberg IX	&	3.60	&	0.24	&	0200980101	&	2004 Sep 26	&	UVW1, UVM2, UVW2	\\
NGC 4395	&	4.00	&	0.05	&	0142830101	&	2003 Nov 30	&	U, B, UVW1, UVW2	\\
NGC 4490	&	7.80	&	0.07	&	0112280201	&	2002 May 27	&	V	\\
$\cdots$	&	$\cdots$	&	$\cdots$	&	0556300201	&	2008 Jun 22	&	U, UVW1, UVM2	\\
NGC 4631	&	7.50	&	0.05	&	0110900201	&	2002 Jun 28	&	U, B, UVW1, UVW2	\\
NGC 4736	&	4.30	&	0.05	&	0094360601	&	2002 May 23	&	V, B	\\
$\cdots$	&	$\cdots$	&	$\cdots$	&	0404980101	&	2006 Nov 27	&	U	\\
NGC 5204	&	4.80	&	0.04	&	0142770101	&	2003 Jan 6	&	UVM2, UVW2	\\
$\cdots$	&	$\cdots$	&	$\cdots$	&	0150650301	&	2003 May 1	&	B	\\
$\cdots$	&	$\cdots$	&	$\cdots$	&	0405690201	&	2006 Nov 19	&	UVW1	\\
M51	&	7.20	&	0.11	&	0303420101	&	2006 May 20	&	U, UVW1	\\
$\cdots$	&	$\cdots$	&	$\cdots$	&	0303420201	&	2006 May 24	&	UVW2	\\
M101	&	7.40	&	0.03	&	0104260101	&	2002 Jun 4	&	UVW1, UVM2, UVW2	\\
$\cdots$	&	$\cdots$	&	$\cdots$	&	0212480201	&	2005 Jan 8	&	V, U, B	\\
$\cdots$	&	$\cdots$	&	$\cdots$	&	0164560701	&	2004 Jul 23	&	U, UVW1, UVM2, UVW2	\\
NGC 5408	&	4.80	&	0.21	&	0500750101	&	2008 Jan 13	&	UVW2	\\
$\cdots$	&	$\cdots$	&	$\cdots$	&	0653380301	&	2010 Jul 19	&	UVM2	\\
NGC 6946	&	5.10	&	1.06	&	0093641701	&	2003 Jun 18	&	UVW1	\\
$\cdots$	&	$\cdots$	&	$\cdots$	&	0200670101	&	2004 Jun 9	&	U	\\
$\cdots$	&	$\cdots$	&	$\cdots$	&	0200670401	&	2004 Jun 25	&	B, UVM2	\\

\enddata
\tablecomments{
(1) Host galaxy;
(2) Galaxy distance adopted from WMR;
(3) Galactic extinction in the direction of each galaxy, see text for details;
(4) XMM-Newton Observation ID;
(5) Observation date;
(6) OM filters used in this paper for each observation;
}
\end{deluxetable}


\clearpage
\pagestyle{empty}  
\begin{deluxetable}{c|rrccccccrrccc}
\tablecolumns{13}   
\tablewidth{0pt}                                                                                   
\tabletypesize{\tiny}  
\rotate                                                                         
\tablenum{2}                                                                                       
\tablecaption{ULX sample and photometric results for the 100~pc regions\label{table2}}          
\tablehead{
\colhead{Src} & \colhead{IAU} & \colhead{ULX} & \colhead{RA} & \colhead{DEC} & \colhead{Exp} & \colhead{L$_X$} & 
\colhead{UVW1} & \colhead{S/N} & \colhead{F$_X$/} & \colhead{Stars}  & \colhead{Literature} & \colhead{Names} & \colhead{Ref} \\

\colhead{} & \colhead{(J2000)} & \colhead{} & \colhead{(J2000)} & \colhead{(J2000)} & \colhead{(Ks)} & \colhead{(erg/s)} & 
\colhead{} & \colhead{} & \colhead{F$_{UVW1}$} & \colhead{} & \colhead{} & \colhead{} & \colhead{} \\

\colhead{(1)} & \colhead{(2)} & \colhead{(3)} & \colhead{(4)} & \colhead{(5)} & \colhead{(6)} & \colhead{(7)} & \colhead{(8)} &
\colhead{(9)} & \colhead{(10)} & \colhead{(11)} & \colhead{(12)} & \colhead{(13)} & \colhead{(14)}\\
}
\startdata

1	&	0047-2047	&	N247 XMM1	&	00 47 03.8 	&	-20 47 46.2	&	4.4	&	39.85	&	17.59	$\pm$	0.04	&	22.6	&	2.10	&	5.2	&	B1-B7 Ib$^1$	&	X1	&	27	\\
2	&	0047-2517	&	N253 XMM1	&	00 47 32.8 	&	-25 17 52.6	&	3.2	&	39.71	&	15.68	$\pm$	0.04	&	20.3	&	1.71	&	43.7	&	$\cdots$	&	PSX-2, X33, C5	&	11, 12, 13	\\
3	&	0047-2520	&	N253 XMM2	&	00 47 22.4 	&	-25 20 55.2	&	3.2	&	39.43	&	17.69	$\pm$	0.07	&	10.7	&	1.68	&	6.8	&	$\cdots$	&	PSX-5, X21, C4	&	11, 12, 13	\\
4	&	0047-2015	&	 N253 XMM6	&	00 47 42.8 	&	-25 15 05.5	&	2.0	&	39.49	&	$>$ 19.09		-	&	$\cdots$	&	$>$ 3.06	&	$<$ O8V	&	$\cdots$	&	X40	&	12	\\
5	&	0318-6636	&	N1313 XMM3 	&	03 18 22.5 	&	-66 36 06.2	&	4.0	&	40.34	&	20.41	$\pm$	0.15	&	5.3	&	3.07	&	{\bf O9V}	&	B2-O7$^2$	&	X-2, U5	&	14, 13	\\
6	&	0345+6804	&	IC342 XMM1 	&	03 45 55.8 	&	+68 04 54.5	&	4.0	&	39.80	&	18.19	$\pm$	0.12	&	4.5	&	2.76	&	4.7	&	F8-G0 Ib$^3$	&	X-1, U6	&	15, 13	\\
7	&	0346+6811	&	IC342 XMM2 	&	03 46 15.0 	&	+68 11 11.2	&	4.0	&	39.93	&	$>$ 18.63		-	&	$\cdots$	&	$>$ 2.10	&	$<$ 3.1	&	$\cdots$	&	X-13	&	15	\\
8	&	0346+6805	&	IC342 XMM3 	&	03 46 48.6 	&	+68 05 43.2	&	4.0	&	40.75	&	8.40	$\pm$	0.00	&	1461.0	&	0.56	&	39026.0	&	$\cdots$	&	X-21	&	15	\\
9	&	0736+6535	&	N2403 XMM1 	&	07 36 25.6 	&	+65 35 40.0	&	4.3	&	39.49	&	19.10	$\pm$	0.05	&	19.3	&	1.74	&	{\bf O3V}	&	OV/BI$^4$	&	X-1, U7	&	16, 13	\\
10	&	0819+7042	&	Ho II XMM1	&	08 19 28.8 	&	+70 42 20.3	&	4.4	&	40.00	&	16.73	$\pm$	0.02	&	134.9	&	1.52	&	8.6	&	B2Ib-O4V$^5$	&	X-1, IXO 31	&	17, 18	\\
11	&	0957+6903	&	Ho IX XMM1 	&	09 57 53.3 	&	+69 03 48.7	&	7.9	&	40.20	&	20.39	$\pm$	0.06	&	16.1	&	2.92	&	{\bf B2III}	&	$>$ B2 (I/III)$^6$	&	Ho X-1, M81 X-9	&	19	\\
12	&	1226+3331	&	N4395 XMM1 	&	12 26 01.5 	&	+33 31 29.0	&	4.3	&	39.43	&	19.33	$\pm$	0.06	&	22.5	&	1.62	&	O3V	&	$\cdots$	&	X-1, IXO 53	&	16, 18	\\
13	&	1230+4139	&	N4490 XMM1 	&	12 30 32.4 	&	+41 39 14.6	&	4.3	&	39.81	&	19.17	$\pm$	0.04	&	45.7	&	1.32	&	7.6	&	$\cdots$	&	ULX4	&	20	\\
14	&	1230+4138	&	N4490 XMM2 	&	12 30 36.5 	&	+41 38 33.3	&	4.3	&	39.67	&	17.84	$\pm$	0.02	&	63.7	&	0.64	&	26.0	&	$\cdots$	&	ULX5	&	20	\\
15	&	1230+4138	&	N4490 XMM3 	&	12 30 43.3 	&	+41 38 11.5	&	4.3	&	40.94	&	19.02	$\pm$	0.04	&	55.1	&	2.39	&	8.8	&	$\cdots$	&	ULX6	&	20	\\
16	&	1230+4139	&	N4490 XMM4 	&	12 30 31.1 	&	+41 39 08.1	&	4.3	&	39.79	&	17.00	$\pm$	0.03	&	43.1	&	1.03	&	56.3	&	$\cdots$	&	ULX3	&	20	\\
17	&	1230+4141	&	N4490 XMM5 	&	12 30 30.3 	&	+41 41 40.3	&	4.3	&	39.47	&	17.78	$\pm$	0.04	&	54.2	&	0.94	&	27.7	&	$\cdots$	&	ULX2	&	20	\\
18	&	1241+3232	&	N4631 XMM1 	&	12 41 55.8 	&	+32 32 14.0	&	2.3	&	39.81	&	16.83	$\pm$	0.03	&	77.9	&	0.73	&	61.4	&	$\cdots$	&	X-1, U29, IXO 65	&	16, 13, 17	\\
19	&	1250+4107	&	N4736 XMM1 	&	12 50 50.2 	&	+41 07 12.0	&	4.4	&	40.25	&	13.69	$\pm$	0.01	&	175.1	&	1.17	&	747.3	&	$\cdots$	&	X-4	&	21	\\
20	&	1329+5825	&	N5204 XMM1 	&	13 29 38.5 	&	+58 25 03.6	&	5.0	&	39.83	&	18.07	$\pm$	0.03	&	76.5	&	1.38	&	8.0	&	B0Ib$^7$	&	X-1, U36, IXO 77	&	16, 13, 17	\\
21	&	1330+4711	&	M51 XMM2 	&	13 30 07.7 	&	+47 11 04.8	&	4.0	&	39.48	&	19.12	$\pm$	0.08	&	13.4	&	1.55	&	6.9	&	F2-5I$^8$	&	ULX 9, IXO 81, 82	&	22, 18, 23	\\
22	&	1329+4710	&	M51 XMM6 	&	13 29 57.5 	&	+47 10 45.3	&	4.0	&	40.54	&	17.34	$\pm$	0.04	&	53.4	&	1.89	&	35.3	&	$\cdots$	&	63	&	23	\\
23	&	1403+5418	&	M101 XMM1 	&	14 03 14.7 	&	+54 18 05.0	&	2.0	&	39.46	&	20.72	$\pm$	0.17	&	5.1	&	1.79	&	O6V	&	$\cdots$	&	U39, XMM-2 	&	13, 24	\\
24	&	1403+5427	&	M101 XMM2 	&	14 03 03.8 	&	+54 27 37.0	&	1.3	&	39.66	&	20.82	$\pm$	0.20	&	4.3	&	2.10	&	O7V	&	$\cdots$	&	U38, XMM-1	&	13, 24	\\
26	&	1404+5426	&	M101 XMM3 	&	14 04 14.6 	&	+54 26 04.4	&	1.3	&	39.57	&	21.04	$\pm$	0.22	&	3.8	&	2.08	&	{\bf B5I}	&	B0I$^9$	&	U41, XMM-3	&	13, 24	\\
25	&	1403-4122	&	N5408 XMM1	&	14 03 19.8	&	-41 22 59.3	&	4.0	&	40.04	&	17.39	$\pm$	0.05	&	37.2	&	1.50	&	8.3	&	$\cdots$	&	X-1	&	25	\\
27	&	2035+6011	&	N6946 XMM1	&	20 35 00.8	&	+60 11 30.6	&	1.9	&	40.00	&	19.14	$\pm$	0.14	&	5.1	&	2.56	&	3.4	&	O/BI$^{10}$	&	X-1, U45, IXO 85	&	26, 13, 17	\\

\enddata
\tablecomments{
NGC 4736 XMM1 and NGC 5408 XMM1 have no images in the UVW1 band.
The values shown for these ULXs were obtained using the U and UVM2 band.
(2) IAU names;
(3) ULX names from WMR;
(6) Exposure time for the UVW1 image used;
(7) The logarithm of the X-ray unabsorbed luminosities for ULXs in the 0.3$-$10 keV band from WMR;    
(8) UVW1 magnitude estimated by integrating the fluxes within the 100~pc regions.
For NGC 253 XMM6 and IC~342 XMM2 we show the 3$\sigma$ lower limit;
(9) Signal to noise ratio in the UVW1 band;
(10) The logarithm of the ratio between the X-ray and UVW1 luminosities;
(11) This column shows estimates for the stars detected within the 100~pc regions.
Assuming the UVW1 emission is due to young stars we estimate the minimum number of O5 V stars 
required to produce an equivalent flux in the UVW1 band.
For the estimates compatible with single stars we show the star types based only on the UVW1 flux.
Four of these we estimate to have actually single star counterparts (Section~3.1), and their types are printed in bold.
(12) The star types for known optical counterparts from the literature;
(13) Common names for ULXs from the literature;
(14) References for the names in col. 13.
}
\tablerefs{
1. \citet{tao12a};
2. \citet{liu07};
3. \citet{feng08};
4. \citet{rob08};
5. \citet{kaa04};
6. \citet{gris06};
7. \citet{liu04};
8. \citet{ter06};
9. \citet{gris12};
10. \citet{kaa10,ber12};
11. \citet{humph03};
12. \citet{vog99};
13. \citet{ber08};
14. \citet{col95};
15. \citet{kong03};
16. \citet{robwar};
17. \citet{leh05};
18. \citet{col02};
19. \citet{mill04};
20. \citet{rob02};
21. \citet{era02};
22. \citet{ter06};
23. \citet{dew05};
24. \citet{jen04};
25. \citet{sor06};
26. \citet{ber12}.
27. \citet{lira00};
}
\end{deluxetable}

\clearpage
\pagestyle{empty}  
\begin{deluxetable}{clrr}
\tablecolumns{3}   
\tablewidth{0pt}                                                                                   
\tabletypesize{\tiny}  
\rotate                                                                         
\tablenum{3}                                                                                       
\tablecaption{ULXs with multiple observations \label{table4}}          
\tablehead{
\colhead{ULX} & \colhead{Date} & \colhead{UT} & \colhead{Mag} \\

\colhead{} & \colhead{} & \colhead{(h)} & \colhead{} \\

\colhead{(1)} & \colhead{(2)} & \colhead{(3)} & \colhead{(4)} \\
}
\startdata

Holmberg II XMM1	&	2002 Apr 16	&	12.171	&	16.735	$\pm$	0.021	\\
$\cdots$	&	2010 Mar 26	&	9.499	&	16.825	$\pm$	0.029	\\
$\cdots$	&	2010 Mar 26	&	11.612	&	16.995	$\pm$	0.017	\\
$\cdots$	&	2010 Mar 26	&	17.837	&	16.875	$\pm$	0.015	\\
IC 342 XMM1	&	2004 Aug 17	&	21.629	&	18.195	$\pm$	0.115	\\
$\cdots$	&	2005 Feb 10	&	20.117	&	18.225	$\pm$	0.117	\\
$\cdots$  &  2001 Feb 11   &    1.345  &      $<$ 18.21 \\
M101 XMM2	&	2004 Jul 23	&	10.751	&	21.037	$\pm$	0.255	\\
$\cdots$	&	2005 Jan 8	&	16.396	&	21.297	$\pm$	0.203	\\
NGC 1313 XMM3	&	2000 Oct 17	&	8.269	&	20.405	$\pm$	0.153	\\
$\cdots$	&	2006 Oct 16	&	13.782	&	20.425	$\pm$	0.111	\\

\enddata
\tablecomments{
The data in the last column is for the UVW1 band, except for M101 XMM2, which is the U band.
}
\end{deluxetable}


\clearpage
\pagestyle{empty}  
\begin{deluxetable}{r|cccc|cccc|cccc|cccc|cccc}
\tablecolumns{20}   
\tablewidth{0pt}                                                                                   
\tabletypesize{\tiny}  
\rotate                                                                         
\tablenum{4}                                                                                       
\tablecaption{Star-forming regions estimates\label{table3}
}          
\tablehead{
\colhead{ULX} & 
\multicolumn{4}{c}{Average Values} & \multicolumn{4}{c}{100 pc regions} & 
\multicolumn{4}{c}{Closest SFR} & \multicolumn{4}{c}{Youngest SFR} & \multicolumn{4}{c}{Most massive SFR} \\
\tableline
\colhead{ULX} & 
\colhead{A$_V$(X)} & \colhead{A$_V$} & \colhead{A$_V$} & \colhead{Age} & 
\colhead{Age} & \colhead{Star} & \colhead{A$_V$} & \colhead{Mass} &
\colhead{Offset} & \colhead{Age} & \colhead{Star} & \colhead{Mass} &
\colhead{Offset} & \colhead{Age} & \colhead{Star} & \colhead{Mass} &
\colhead{Offset} & \colhead{Age} & \colhead{Star} & \colhead{Mass} \\
\colhead{} & 
\colhead{} & \colhead{Range} & \colhead{Avg.} & \colhead{Myr} &
\colhead{Myr} & \colhead{} & \colhead{} & \colhead{log M$_{\sun}$} &
\colhead{pc} & \colhead{Myr} & \colhead{} & \colhead{log M$_{\sun}$} &
\colhead{pc} & \colhead{Myr} & \colhead{} & \colhead{log M$_{\sun}$} &
\colhead{pc} & \colhead{Myr} & \colhead{} & \colhead{log M$_{\sun}$} \\
\colhead{(1)} & 
\colhead{(2)} & \colhead{(3)} & \colhead{(4)} & 
\colhead{(5)} & \colhead{(6)} & \colhead{(7)} & \colhead{(8)} & 
\colhead{(9)} & \colhead{(10)} & \colhead{(11)} & \colhead{(12)} & 
\colhead{(13)} & \colhead{(14)} & \colhead{(15)} & \colhead{(16)} 
& \colhead{(17)} & \colhead{(18)} & \colhead{(19)} & \colhead{(20)} & \colhead{(21)}\\
}
\startdata

N247 XMM1	&	2.31	&	0$-$1.5	&	0.83	&	18	&	10	&	B0.5	&	1.0	&	3.8	&	64	&	10	&	B0.5	&	3.8	&	492	&	3	&	O6	&	3.8	&	190	&	10	&	B0.5	&	5.2	\\
N253 XMM1	&	2.82	&	1$-$2.5	&	1.75	&	500	&	900	&	A5	&	1.0	&	6.3	&	211	&	100	&	B4.5	&	6.9	&	309	&	100	&	B4.5	&	6.9	&	309	&	100	&	B4.5	&	6.9	\\
N253 XMM2	&	1.01	&	1.00	&	1.00	&	13	&	100	&	B4.5	&	1.0	&	4.8	&	60	&	8	&	B0	&	4.1	&	60	&	8	&	B0	&	4.1	&	462	&	15	&	B1.5	&	4.7	\\
 N253 XMM6	&	3.55	&	1$-$1.2	&	0.73	&	118	&	$\cdots$	&	$\cdots$	&	$\cdots$	&	$<$3.4	&	{\bf 221}	&	5	&	O9	&	4.6	&	221	&	5	&	O9	&	4.6	&	221	&	150	&	B5.5	&	5.1	\\
N1313 XMM3 	&	3.49	&	0.50	&	0.50	&	5	&	$\cdots$	&	$\cdots$	&	$\cdots$	&	3.1	&	68	&	$\cdots$	&	$\cdots$	&	3.2	&	106	&	5	&	O9	&	3.0	&	222	&	$\cdots$	&	$\cdots$	&	3.6	\\
IC342 XMM1 	&	3.27	&	$\cdots$	&	$\cdots$	&	$\cdots$	&	$\cdots$	&	$\cdots$	&	$\cdots$	&	3.0	&	$\cdots$	&	$\cdots$	&	$\cdots$	&	$\cdots$	&	$\cdots$	&	$\cdots$	&	$\cdots$	&	$\cdots$	&	$\cdots$	&	$\cdots$	&	$\cdots$	&	$\cdots$	\\
IC342 XMM2 	&	13.47	&	$\cdots$	&	$\cdots$	&	$\cdots$	&	$\cdots$	&	$\cdots$	&	$\cdots$	&	$<$2.8	&	$\cdots$	&	$\cdots$	&	$\cdots$	&	$\cdots$	&	$\cdots$	&	$\cdots$	&	$\cdots$	&	$\cdots$	&	$\cdots$	&	$\cdots$	&	$\cdots$	&	$\cdots$	\\
IC342 XMM3 	&	5.47	&	2.75	&	2.75	&	5	&	10	&	B0.5	&	2.0	&	7.0	&	{\bf 33}	&	5	&	O9	&	7.1	&	40	&	5	&	O9	&	7.0	&	33	&	5	&	O9	&	7.1	\\
N2403 XMM1 	&	1.30	&	0$-$0.5	&	0.20	&	151	&	$\cdots$	&	$\cdots$	&	$\cdots$	&	3.7	&	103	&	100	&	B4.5	&	4.3	&	361	&	5	&	O9	&	4.5	&	132	&	400	&	A0	&	5.4	\\
Ho II XMM1	&	0.87	&	$\cdots$	&	$\cdots$	&	$\cdots$	&	$\cdots$	&	$\cdots$	&	$\cdots$	&	4.5	&	{\bf 83}	&	$\cdots$	&	$\cdots$	&	5.1	&	$\cdots$	&	$\cdots$	&	$\cdots$	&	$\cdots$	&	83	&	$\cdots$	&	$\cdots$	&	5.1	\\
Ho IX XMM1 	&	1.18	&	$\cdots$	&	$\cdots$	&	$\cdots$	&	$<$400	&	A0	&	$<$1.0	&	3.2	&	{\bf 25}	&	$\cdots$	&	$\cdots$	&	3.8	&	$\cdots$	&	$\cdots$	&	$\cdots$	&	$\cdots$	&	25	&	$\cdots$	&	$\cdots$	&	3.8	\\
N4395 XMM1 	&	1.13	&	0$-$0.5	&	0.30	&	19	&	5	&	B4.5	&	0.5	&	2.8	&	{\bf 28}	&	50	&	B3.5	&	4.1	&	443	&	5	&	O9	&	4.2	&	240	&	10	&	B0.5	&	4.3	\\
N4490 XMM1 	&	3.27	&	$\cdots$	&	$\cdots$	&	$<$ 180	&	30	&	B2	&	1.0	&	5.0	&	171	&	$<$ 180	&	B6.5	&	$<$ 5.5	&	418	&	$<$ 50	&	B3.5	&	$<$ 5.3	&	364	&	$<$ 150	&	B5.5	&	$<$ 5.5	\\
N4490 XMM2 	&	2.48	&	$\cdots$	&	$\cdots$	&	$\cdots$	&	12	&	B0.5	&	0.5	&	4.8	&	{\bf 116}	&	$\cdots$	&	$\cdots$	&	6.2	&	$\cdots$	&	$\cdots$	&	$\cdots$	&	$\cdots$	&	116	&	$\cdots$	&	$\cdots$	&	6.2	\\
N4490 XMM3 	&	7.33	&	$\cdots$	&	$\cdots$	&	$<$ 75	&	3	&	O6	&	2.5	&	4.9	&	146	&	$<$ 75	&	B3.5	&	5.5	&	146	&	$<$ 75	&	B.3.5	&	5.5	&	146	&	$<$ 75	&	B.3.5	&	5.5	\\
N4490 XMM4 	&	5.75	&	0$-$1.5	&	0.88	&	5	&	10	&	B0.5	&	0.5	&	4.4	&	133	&	15	&	B1.5	&	4.9	&	192	&	1	&	O5	&	4.6	&	507	&	4	&	O8.5	&	5.6	\\
N4490 XMM5 	&	2.20	&	0.5$-$1.0	&	0.77	&	2	&	12	&	B0.5	&	0.5	&	4.3	&	{\bf 87}	&	$<$ 50	&	$\cdots$	&	$<$ 5.1	&	452	&	1	&	O5	&	4.5	&	647	&	1	&	O5	&	5.2	\\
N4631 XMM1 	&	1.69	&	0.50	&	0.50	&	7	&	10	&	B0.5	&	0.5	&	4.7	&	{\bf 69}	&	5	&	O9	&	5.3	&	235	&	5	&	O9	&	5.0	&	671	&	5	&	O9	&	5.3	\\
N4736 XMM1 	&	3.55	&	2.00	&	2.00	&	8	&	15	&	B1.5	&	2.0	&	5.8	&	246	&	3	&	O6	&	6.0	&	246	&	3	&	O6	&	6.0	&	274	&	10	&	B0.5	&	7.0	\\
N5204 XMM1 	&	0.50	&	0$-$0.5	&	0.15	&	20	&	15	&	B1.5	&	0.0	&	3.9	&	{\bf 52}	&	30	&	B2	&	4.7	&	309	&	3	&	O6	&	3.9	&	321	&	18	&	B1.5	&	4.8	\\
M51 XMM2 	&	0.73	&	0.5$-$1.0	&	0.75	&	12	&	$\cdots$	&	$\cdots$	&	$\cdots$	&	3.9	&	{\bf 94}	&	20	&	B1.5	&	4.7	&	453	&	3	&	O6	&	3.7	&	354	&	$\cdots$	&	$\cdots$	&	4.7	\\
M51 XMM6 	&	4.62	&	0.5$-$1.25	&	0.75	&	48	&	$\cdots$	&	$\cdots$	&	$\cdots$	&	4.6	&	157	&	5	&	O9	&	4.7	&	157	&	5	&	87.1	&	4.7	&	229	&	150	&	B5.5	&	5.8	\\
M101 XMM1 	&	0.12	&	0$-$0.75	&	0.42	&	11	&	$\cdots$	&	$\cdots$	&	$\cdots$	&	3.5	&	{\bf 74}	&	15	&	B1.5	&	4.0	&	484	&	5	&	O9	&	4.6	&	708	&	10	&	B0.5	&	5.7	\\
M101 XMM2 	&	0.90	&	0$-$1.0	&	0.38	&	56	&	3	&	O6	&	1.0	&	2.8	&	{\bf 84}	&	5	&	O9	&	4.0	&	812	&	3	&	O6	&	3.3	&	178	&	300	&	B8	&	4.6	\\
M101 XMM3	&	1.12	&	0$-$0.5	&	0.35	&	206	&	$\cdots$	&	$\cdots$	&	$\cdots$	&	3.5	&	678	&	3	&	O6	&	3.8	&	678	&	3	&	O6	&	3.8	&	930	&	1000	&	F0	&	7.7	\\
N5408 XMM1	&	0.51	&	$\cdots$	&	$\cdots$	&	$\cdots$	&	$\cdots$	&	$\cdots$	&	$\cdots$	&	4.6	&	106	&	$\cdots$	&	$\cdots$	&	5.2	&	$\cdots$	&	$\cdots$	&	$\cdots$	&	$\cdots$	&	401	&	$\cdots$	&	$\cdots$	&	6.1	\\
N6946 XMM1	&	1.87	&	1.00	&	1.00	&	12	&	$\cdots$	&	$\cdots$	&	$\cdots$	&	3.4	&	{\bf 30}	&	$\cdots$	&	$\cdots$	&	3.9	&	225	&	12	&	B0.5	&	4.6	&	225	&	12	&	B0.5	&	4.6	\\

\enddata
\tablecomments{
(2) Extinction estimated using the column densities obtained by WMR from X-ray spectral fits;
(3) Range of extinction values for star-forming regions around each ULX;
(4) - (5) Average extinction and cluster age for the star-forming regions around each ULX;
(6) - (9) Results obtained for the 100~pc regions by comparing the OM colors with Starburst99 tracks: 
cluster age, the earliest stars that survive in the cluster, given its age, and the mass of the cluster;
(10) - (13) Same results for the closest SFR for each ULX, plus the offset distance to the center of the SFR. 
Some ULXs are actually located inside these SFR, the offset distances for these are shown in bold.
(14) - (17) Same results for the youngest SFR around each ULX, plus the offset distance to the center of the SFR. 
(18) - (21) Same results for the closest SFR for each ULX, plus the offset distance to the center of the SFR. 
The final row shows median values for each column.
}
\end{deluxetable}


\clearpage

\end{document}